%% file: iclr2027_conference.tex
\PassOptionsToPackage{table}{xcolor}
\documentclass{article} % For LaTeX2e
\usepackage{iclr2027_conference,times}

\input{math_commands.tex}

\usepackage{hyperref}
\usepackage{url}
\usepackage{graphicx}
\usepackage{booktabs}
\usepackage{amsmath}
\usepackage{amssymb}
\usepackage{subcaption}
\usepackage{multirow}
\usepackage{float}
\usepackage{xcolor}
\usepackage{tikz}
\usetikzlibrary{arrows.meta,positioning,fit,calc}

\newcommand{\best}[1]{\textbf{#1}}

\newcommand{\NFE}{\mathrm{NFE}}
\title{Where Should Physics Enter a Molecular Crystal Generator?}

\author{
Haocheng Tang \\
Khoury College of Computer Science\\
Northeastern University\\
Boston, MA 02115, USA
\And
Junmei Wang\thanks{Corresponding authors: 
\texttt{juw79@pitt.edu} and \texttt{w.jin@northeastern.edu}.} \\
School of Pharmacy\\
University of Pittsburgh\\
Pittsburgh, PA 15213, USA
\And
Wengong Jin\textsuperscript{*} \\
Khoury College of Computer Science\\
Northeastern University\\
Boston, MA 02115, USA
}

\iclrfinalcopy % Uncomment for camera-ready version, but NOT for submission.

\begin{document}

\maketitle

\begin{abstract}
Generative models make molecular crystal structure prediction fast, but their samples still exhibit geometric and packing violations. Physics can be introduced during training, post-training, or inference, yet these choices are rarely compared with the generator and physical signal held fixed. We introduce CrystAF, an all-atom crystal flow-map generation model, and use it with the UMA interatomic potential to systematically study where physics should enter. Post-training learns physical preferences directly into CrystAF, improving molecular validity and crystal packing while leaving sampling unchanged: physics is paid for once during training rather than repeatedly at deployment. In contrast, UMA relaxation is effective at repairing local clashes but makes generation 6--26$\times$ slower, while learning from relaxed targets provides little benefit. These routes are complementary rather than competing. Physics-informed post-training first shifts the generated distribution toward more physically reasonable structures, after which inexpensive inference-time corrections further remove clashes and restore stereochemistry that the generator cannot represent. Importantly, the same post-training strategy also improves the multi-step all-atom Clari-M and rigid-body MolCrystalFlow generators, demonstrating transfer across architectures and representations. Together, our results suggest a simple principle: learn reusable physical alignment into the generator, and reserve inference-time physics for residual constraints that are better corrected than learned.

\end{abstract}

\section{Introduction}

The same organic molecule can crystallize in several forms, and the difference can be costly. A famous story is the oral HIV drug: ritonavir. Two years after the molecule reached the market, a more stable polymorph appeared, resulting in treatment failure. The drug had to be withdrawn \citep{bauer2001ritonavir,morissette2003ritonavir}. Crystal structure prediction (CSP) tries to find such forms computationally, by proposing many candidate packings for a molecule and ranking them by energy \citep{price2014csp,reilly2016blindtest}. Generative models are now a fast source of these proposals. Clari \citep{lo2026clari} and MolCrystalFlow \citep{zeng2026molcrystalflow} learn packings from the Cambridge Structural Database (CSD) and produce plausible unit cells in seconds.

What these models learn, though, is the data, not the physics. A sample can be likely under the model and still contain overlapping atoms, a strained molecule, or a cell that is too large. So physics is added somewhere, in one of three places. At \emph{training time}, the model can be supervised with physically refined structures or physics-based losses \citep{noe2019boltzmann}. In \emph{post-training}, an energy model becomes a reward for reinforcement learning \citep{black2024ddpo,fan2023dpok,liu2025flowgrpo,zheng2026diffusionnft}, as in PackFlow with a machine-learned interatomic potential (MLIP) \citep{subramanian2026packflow} and in \citet{wang2026controllable} for 3D molecules. At \emph{inference time}, no training is needed: each sample is steered by energy or force guidance \citep{dhariwal2021diffusion,chung2023dps,bao2023eegsde}, relaxed with MMFF \citep{halgren1996mmff} or an MLIP, or corrected by parity inversion \citep{hassan2024etflow,nikitin2026loqi} or constraint projection \citep{cai2026pcfm}. Each option has a catch. Inference-time methods come with hand-set physical hyper-parameters, and because the physics never enters the generator, the energy model must be called again for each run. Post-training pays that cost once during training, but if the reward resembles the evaluation metric, a higher score may mean only that the model has learned the metric \citep{gao2023overoptimization}. Training-time supervision needs a physical target, and it often relates to model architecture change.

\begin{figure}[t]
\centering
\includegraphics[width=\textwidth]{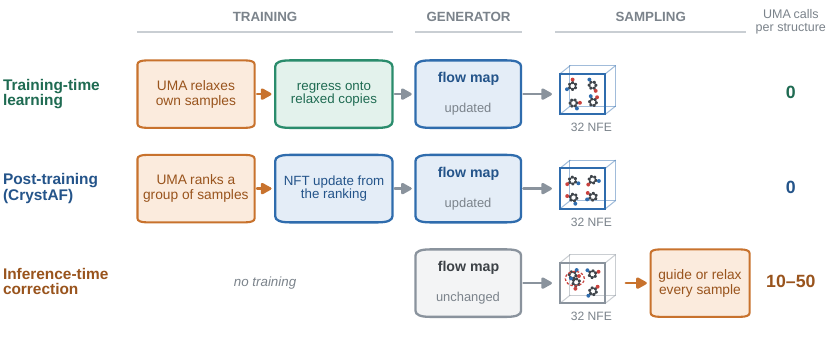}
\caption{Three ways to inject the same UMA physical signal into CrystAF. Training-time and post-training methods use UMA during training; inference-time correction applies it to each sample. PoseBusters, PDD, and COMPACK are used only for evaluation.}
\label{fig:overview}
\end{figure}

Three questions follow. Where should physics enter: into the weights or into each sample? When a metric improves, is the model genuinely more physical, or merely better optimized for that metric? And which physical constraints can be learned by the generator, versus only enforced at inference time when they lie outside its representation, as with stereochemistry? Published methods cannot answer these questions cleanly because they differ in generator, physical signal, and evaluation. We isolate these choices with everything else fixed (Figure~\ref{fig:overview}). CrystAF is an all-atom crystal flow map distilled from Clari for few-step sampling, and UMA \citep{wood2025uma} provides the shared physical signal through single-point energies, forces, and stresses. Only the entry point changes: UMA supplies regression targets, a post-training reward, or inference-time forces. We evaluate with PoseBusters \citep{buttenschoen2024posebusters}, pointwise-distance distributions \citep{widdowson2022amd}, and COMPACK matches \citep{motherwell2025compack}, none of which is optimized directly. The 32-NFE setting also makes inference cost consequential: dozens of extra potential calls can dominate generation time.

We train CrystAF once and compare all three routes, adapting negative-aware fine-tuning \citep{zheng2026diffusionnft} to the flow map for post-training. Five findings stand out. Post-training improves validity and packing at no sampling cost, raising PoseBusters from $89.3\%$ to $92.0\%$, reducing clashes from $11.3\%$ to $9.6\%$, and improving cell volume and distance distribution. Ten-step UMA relaxation halves clashes at a similar pass rate, but leaves the cell unchanged and makes sampling six times slower. Simple rejection explains most of the validity gain, while UMA ranking adds packing improvements. The update rule matters as much as the reward: plain DiffusionNFT reaches a similar pass rate but worsens cell volume and distance distribution. The routes also stack: inference-time corrections reduce clashes to $2.8\%$ and raise stereochemical agreement from $48\%$ to $95\%$, enforcing constraints such as chirality that the generator cannot represent. Finally, post-training improves physical quality without consistently changing packing recovery, whereas inference-time corrections improve recovery by turning near-misses into valid matches. Our contributions are:
\begin{itemize}
\item \textbf{Controlled comparison of where physics enters.}
We compare training-time, post-training, and inference-time physics with the generator and physical signal fixed, measuring both crystal quality and inference cost.

\item \textbf{A fast, physics-alignable crystal flow map.}
We introduce CrystAF, a few-step all-atom crystal flow map, together with a post-training recipe that converts black-box MLIP feedback into within-family advantages for negative-aware fine-tuning.

\item \textbf{An analysis of what physical alignment can and cannot learn.}
We evaluate gains on metrics never optimized directly, ablate both the physical signal and the update rule, and separate properties that can be learned through post-training from constraints that must be enforced explicitly at inference time.
\end{itemize}

\section{Background and related work}
\label{sec:background}

\paragraph{Generative crystal structure prediction.}
Diffusion and flow models now generate periodic crystals directly, from inorganic materials \citep{xie2022cdvae,jiao2023diffcsp,zeni2025mattergen,li2025pccd} to organic molecular crystals. MolCrystalFlow moves rigid molecules inside a cell \citep{zeng2026molcrystalflow}. Clari generates every atom of the unit cell with a diffusion transformer, conditioned on the molecular graph and the number of molecules per cell \citep{lo2026clari}. We use Clari's state $x=[\tfrac12 L;\,C]/\sigma$, which stacks the cell matrix $L\in\mathbb R^{3\times3}$ and the Cartesian coordinates $C$ and divides them by a fixed scale $\sigma$. Time runs from noise at $t=0$ to data at $t=1$.

\paragraph{Few-step flow maps and post-training.}
An ordinary flow model predicts a velocity $v(z_s,s)$ and must integrate it in many small steps. A flow map instead predicts the \emph{average} velocity between two times $r<t$ \citep{geng2025meanflow}:
\begin{equation}
U(z_r,r,t)=\frac{1}{t-r}\int_r^t v(z_s,s)\,ds,\qquad
z_t=z_r+(t-r)U(z_r,r,t).
\label{eq:flow-map-definition}
\end{equation}
The second identity is the point: knowing $U$ lets the sampler jump directly from $z_r$ to $z_t$, so a jump of any length costs one network evaluation (NFE). AnyFlow builds $U$ as a frozen teacher velocity plus a learned correction \citep{gu2026anyflow}.

Pretrained diffusion and flow models can be post-trained with forward-process methods such as DiffusionNFT \citep{zheng2026diffusionnft}, which use sample-level rewards to shift the generator toward preferred outputs. CrystAF is slightly different because it predicts an interval-average flow map rather than an instantaneous velocity. We therefore follow MeanFlowNFT \citep{huang2026meanflownft} and recover the induced instantaneous velocity before applying the NFT update. The correction depends on the jump length, which becomes important in the few-step regime (Section~\ref{sec:postrain}).

\paragraph{Inference-time physical correction.}
Rather than changing model weights, physics can be applied directly to each generated sample. Energy- or force-based guidance steers the sampling trajectory \citep{dhariwal2021diffusion, chung2023dps, bao2023eegsde}, while post-sampling relaxation refines completed structures with a force field or MLIP \citep{halgren1996mmff, wood2025uma}. Constraint-based methods such as PCFM project samples toward feasible molecular geometry \citep{cai2026pcfm}, while explicit geometric corrections can restore properties such as stereochemistry \citep{hassan2024etflow, nikitin2026loqi}. These methods require no retraining and can enforce properties absent from the generator's representation, but potential-based variants incur repeated inference cost and introduce hand-set step sizes, force scales, or iteration counts.

\section{Three routes for one physical signal}
\label{sec:method}

\subsection{The generator: CrystAF}
\label{sec:generator}

CrystAF is an all-atom flow map distilled from a Clari-M teacher with AnyFlow (Figure~\ref{fig:arch} in the appendix). CrystAF-base initializes every experiment. A frozen Clari time embedding reads the start time $r$, while a trainable copy and rank-32 attention LoRA read the target time $t$; only $3.8\%$ of the parameters are trainable. One adapter supports $8$--$50$ NFE on $t_i=(i/N)^\rho$; the main text uses $N=32$ ($\rho=1$), versus 99 NFE for Clari. Distillation details are in Appendix~\ref{sec:si_distillation}.

\subsection{The physical signal, and what it does not see}
\label{sec:signal}

For candidate $i$ of molecular family $g$, UMA-OMC returns energy per molecule $E_{gi}$, mean and maximum atomic force $\bar F_{gi}$ and $F^{\max}_{gi}$, and cell stress $\sigma_{gi}$ \citep{wood2025uma}. A candidate is \emph{eligible} if its minimum interatomic distance exceeds $0.8$\,\AA, its density lies in $[0.3,3]$\,g\,cm$^{-3}$, and UMA returns a finite energy. Post-training may additionally use relative cell-volume error. Our evaluation deliberately uses signals outside this reward. PoseBusters measures molecular geometry and steric validity, pointwise distance distributions (PDD) measures agreement of the full interatomic-distance distribution with the experimental crystal, and COMPACK tests whether the experimental packing is recovered at all. None is directly optimized by UMA; their overlap with individual signal terms is summarized in Table~\ref{tab:overlap}.

\subsection{Post-training CrystAF with UMA feedback}
\label{sec:postrain}

\begin{figure}[t]
\centering
\includegraphics[width=\textwidth]{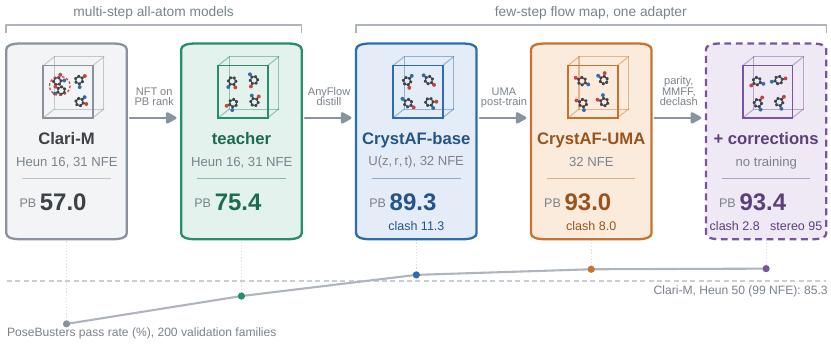}
\caption{Building CrystAF from Clari-M and applying physics to the resulting 32-NFE sampler. Boxes report PoseBusters pass rate on 200 validation families. Clari-M and the teacher use 16 Heun steps; CrystAF uses 32 jumps at a comparable NFE budget. UMA post-training changes the weights, whereas the inference-time chain corrects each sample.}
\label{fig:crystaf}
\end{figure}

Figure~\ref{fig:crystaf} summarizes the CrystAF pipeline and the two physics routes applied after distillation.

\paragraph{From scores to advantages.}
For each training family $g$, we draw $K=100$ candidates from the current model and score them with UMA. Let $m_{gi}\in\{0,1\}$ indicate whether candidate $i$ is eligible. Each channel $c\in\{E,\bar F,F_{\max},\sigma,V\}$ is oriented so that larger is better and standardized within the family as family-standardized channel score $z^c_{gi}$, where $V$ denotes the optional volume-error channel. For the UMA channels, ineligible candidates are assigned the valid-candidate mean before standardization so that failed scores do not distort the group statistics. The advantage is
\begin{equation}
A_{gi} = \operatorname{clip}_{[-1,1]} \left(S_{gi} - \operatorname{med}_{j:\,m_{gj}=1} S_{gj} \right),
\qquad
S_{gi} = \lambda z^{E}_{gi} + (1-\lambda)z^{\bar F}_{gi} + z^{F_{\max}}_{gi} + z^{\sigma}_{gi} + w_V z^{V}_{gi}.
\label{eq:advantage}
\end{equation}
We additionally require $A_{gi}\leq 0$ whenever $m_{gi}=0$. We use $\lambda=0.5$ to balance energy and mean force and $w_V=1$ for the optional volume term. This gate is essential: without it, an invalid structure can receive positive advantage from another channel (Appendix~\ref{sec:rewardbugs}). Full construction details are given in Appendix~\ref{sec:objective-details}.

\paragraph{Negative-aware fine-tuning.}
We map the advantage to $p_{gi}=(A_{gi}+1)/2\in[0,1]$ and apply DiffusionNFT \citep{zheng2026diffusionnft}. For a re-noised candidate $x_r=(1-r)\epsilon+rx_1$, define
$V^{+}_\theta=(1-\beta)V_{\mathrm{old}}+\beta V_\theta$ and
$V^{-}_\theta=(1+\beta)V_{\mathrm{old}}-\beta V_\theta$. We optimize
\begin{equation}
\mathcal L=\mathbb E\Big[\tfrac{p_{gi}}{\beta}\,\ell\big(x_r+(1-r)V^{+}_\theta,\,x_1\big)+\tfrac{1-p_{gi}}{\beta}\,\ell\big(x_r+(1-r)V^{-}_\theta,\,x_1\big)\Big]+\lambda_{\mathrm{KL}}\,\mathbb E\,\lVert V_\theta-V_{\mathrm{ref}}\rVert^2 .
\label{eq:main-nft-objective}
\end{equation}
Here $\ell$ is a per-sample normalized squared error, $\beta=0.1$, and the final term anchors the policy to $V_{\mathrm{ref}}$. UMA affects training only through the advantages; no gradient passes through the scorer.

\paragraph{What changes for a flow map.}
Equation~\ref{eq:main-nft-objective} requires an instantaneous velocity, whereas CrystAF predicts the interval average $U$. Following the flow-map identity,
\begin{equation}
V(x_r,r)=U(x_r,r,t)-(t-r)\,D_rU(x_r,r,t),
\label{eq:identity}
\end{equation}
where $D_rU$ moves $(x_r,r)$ jointly along the flow (Appendix~\ref{sec:flow-derivation}). Because the correction scales with $t-r$, training on mostly short or zero-length intervals suppresses the flow-map-specific update. MeanFlowNFT's default mix yields only $0.75\%$ gradient difference from plain DiffusionNFT; we instead sample adjacent knots of an $N{=}8$ grid, raising it to $3$--$24\%$ across the tested jump lengths (Figure~\ref{fig:dose}). We estimate $D_rU$ by a central difference on the old policy and share it across all three velocity branches.

Training rollouts use 16 NFE, and CrystAF-base begins to drift after several epochs. We therefore fix epoch 4 for the main text and epoch 6 for the appendix; four epochs of 20 families cost about 5 L40S GPU-hours (Appendices~\ref{sec:grid} and~\ref{sec:train-signals}).

\subsection{Training-time learning: relax and distill}
\label{sec:distill-train}

The training-time baseline uses UMA to construct targets rather than rewards. For each family, we draw 24 candidates, apply 25 fixed-cell UMA relaxation steps, and regress the flow map onto the relaxed structure $x^\star$:
\begin{equation}
\mathcal L_{\mathrm{RD}}=\mathbb E\,\big\lVert U_\theta(x_r,r,t)-(x^\star-\epsilon)\big\rVert^2,\qquad
x_r=(1-r)\epsilon+r\,x^\star .
\end{equation}
Because the path from $\epsilon$ to $x^\star$ is linear, $x^\star-\epsilon$ serves as the target for every $(r,t)$. Initialization, families, and epoch budget match the post-training baseline.

\subsection{Inference-time correction: training-free baselines}
\label{sec:trainfree}

\textbf{UMA force guidance.}
Guidance corrects the model's guess of where a sample will end up. At a jump from $r$ to $t$ (for $r\ge r_0$), the flow map predicts the endpoint $\hat x_1=x_r+(1-r)U$. We move its atoms along the UMA forces, $\hat x_1'=\hat x_1+\operatorname{cap}_{\delta}(\eta F(\hat x_1))$, where the cap limits any atom's displacement to $\delta$. The step is then aimed at the corrected endpoint, $x_t=x_r+\frac{t-r}{1-r}(\hat x_1'-x_r)$. Early jumps therefore apply only a fraction $\frac{t-r}{1-r}$ of the correction, and the last jump applies all of it; this is the same slot that PCFM uses for constraint projection \citep{cai2026pcfm}. Each guided jump costs one UMA call. There are three knobs: the force scale $\eta$, the cap $\delta$, and the start time $r_0$

\textbf{UMA relaxation.}
After sampling, we take $K$ fixed-cell steepest-descent steps on UMA forces with the cell held fixed, each capped at $0.05$\,\AA{} maximum atomic displacement. This costs $K$ UMA calls per structure. We keep the cell fixed because UMA cell relaxation worsens agreement with experimental volumes (Appendix~\ref{sec:trainfree-details}).

\textbf{Constraint projection without chirality.}
Following PCFM \citep{cai2026pcfm}, we project finished samples onto RDKit distance-geometry constraints using six closed-form Gauss--Newton iterations, capped at $0.25$\,\AA{} per atom. We omit its chirality term to isolate geometric projection from stereochemical correction.

\textbf{Classical corrections.}
A potential-free chain applies parity inversion to match input R/S labels, restrained MMFF94 relaxation \citep{halgren1996mmff} at most $0.15$\,\AA each atom, and rigid-body declashing. An optional cell calibration multiplies the volume by $0.985$, fitted on the training split.

\section{Experiments}
\label{sec:results}

\textbf{Protocol and Baselines}
We evaluate on the first 200 families of Clari's validation split, 20 samples each, with Clari's metrics: PoseBusters pass rate (PB), steric clash rate, relative cell-volume error, and the earth mover's distance between pointwise distance distributions (PDD). Repeating a configuration moves PB by about $\pm0.5$, so smaller differences count as ties. Every matched post-trained arm starts from CrystAF-base and is read out at epoch 4; no arm gets its own checkpoint selection. The released CrystAF-UMA model, trained for six epochs from a warm-started checkpoint, is shown for reference. Cost is the wall-clock time to generate one structure on one L40S GPU, correction included, together with the number of UMA calls it needs (Appendix~\ref{sec:cost}). \emph{Generator}: CrystAF-base and Clari-M. \emph{Algorithm}: DiffusionNFT applied to $U$ with the same reward, which skips Equation~\ref{eq:identity}; Flow-GRPO appears in Section~\ref{sec:generality}. \emph{Signal}: energy only; energy, forces, and stress; feasibility only; the eligibility gate alone; and a PoseBusters-ranked reward, which optimizes the metric directly. \emph{Injection}: the methods of Sections~\ref{sec:distill-train} and~\ref{sec:trainfree}.

\subsection{What each route fixes, and what it costs}
\label{sec:main}

% Preamble:
% \usepackage[table]{xcolor}

\begin{table}[t]
\caption{Controlled comparison of physics-injection routes. Means $\pm$ SE over 200 validation families $\times$ 20 samples. CrystAF uses 32 NFE; Clari-M uses 99 NFE. ``UMA'' denotes potential calls per sample at inference; ``ms'' is wall-clock time per sample on one L40S; ``GPU-h'' is training cost beyond CrystAF-base. PB and clash are percentages; Stereo reports R/S agreement only when stereochemistry is explicitly enforced. \textbf{Bold} marks the best value per column.}
\label{tab:main}
\centering
\small
\setlength{\tabcolsep}{2.2pt}

\resizebox{\textwidth}{!}{%
\begin{tabular}{lrrrccccc}
\toprule
& \multicolumn{3}{c}{Cost} & \multicolumn{5}{c}{Quality} \\
\cmidrule(lr){2-4}\cmidrule(lr){5-9}
Method & UMA & ms & GPU-h & PB $\uparrow$ & Clash $\downarrow$ & Vol.Err $\downarrow$ & PDD $\downarrow$ & Stereo $\uparrow$ \\
\midrule

\rowcolor{gray!10}
\multicolumn{9}{c}{\textit{No physics}} \\
Clari-M & 0 & 289 & --- & 85.30 \small{$\pm 0.80$} & 9.05 \small{$\pm 0.74$} & 1.83 \small{$\pm 0.13$} & 10.19 \small{$\pm 0.36$} & -- \\
CrystAF-base & 0 & 396 & --- & 89.32 \small{$\pm 0.67$} & 11.33 \small{$\pm 0.84$} & 1.95 \small{$\pm 0.09$} & 10.44 \small{$\pm 0.16$} & -- \\

\midrule
\rowcolor{gray!10}
\multicolumn{9}{c}{\textit{Training-time learning}} \\
Relax-and-distill & 0 & 396 & 6 & 88.20 \small{$\pm 0.68$} & 11.08 \small{$\pm 0.83$} & 1.74 \small{$\pm 0.09$} & 10.21 \small{$\pm 0.18$} & -- \\

\midrule
\rowcolor{gray!10}
\multicolumn{9}{c}{\textit{Post-training alignment}} \\
CrystAF-UMA (matched) & 0 & 398 & 5 & 92.02 \small{$\pm 0.44$} & 9.64 \small{$\pm 0.75$} & \best{1.69} \small{$\pm 0.09$} & \best{10.02} \small{$\pm 0.14$} & -- \\
CrystAF-UMA (released) & 0 & 398 & 8 & 93.01 \small{$\pm 0.43$} & 7.98 \small{$\pm 0.70$} & 1.88 \small{$\pm 0.10$} & 10.25 \small{$\pm 0.20$} & -- \\

\midrule
\rowcolor{gray!10}
\multicolumn{9}{c}{\textit{Inference-time correction of CrystAF-base}} \\
+ UMA guidance, gentle & 16 & 3752 & --- & 91.23 \small{$\pm 0.57$} & 9.45 \small{$\pm 0.75$} & 1.94 \small{$\pm 0.10$} & 10.42 \small{$\pm 0.17$} & -- \\
+ UMA guidance, strong & 16 & 3812 & --- & 90.40 \small{$\pm 0.56$} & 6.85 \small{$\pm 0.62$} & 1.97 \small{$\pm 0.10$} & 10.53 \small{$\pm 0.18$} & -- \\
+ UMA relaxation (10) & 10 & 2462 & --- & 92.90 \small{$\pm 0.43$} & 4.97 \small{$\pm 0.48$} & 1.95 \small{$\pm 0.11$} & 10.51 \small{$\pm 0.23$} & -- \\
+ UMA relaxation (50) & 50 & 10454 & --- & 92.68 \small{$\pm 0.46$} & 5.42 \small{$\pm 0.47$} & 2.01 \small{$\pm 0.09$} & 10.44 \small{$\pm 0.16$} & -- \\
+ PCFM projection & 0 & 396 & --- & 89.43 \small{$\pm 0.65$} & 10.63 \small{$\pm 0.80$} & 1.94 \small{$\pm 0.10$} & 10.28 \small{$\pm 0.19$} & -- \\
+ parity, MMFF, declash & 0 & 676 & --- & 92.91 \small{$\pm 0.41$} & \best{2.64} \small{$\pm 0.46$} & 1.89 \small{$\pm 0.10$} & 10.64 \small{$\pm 0.17$} & 94.9 \\

\midrule
\rowcolor{gray!10}
\multicolumn{9}{c}{\textit{Inference-time correction of CrystAF-UMA}} \\
+ UMA guidance, gentle & 16 & 3752 & 8 & \best{93.78} \small{$\pm 0.35$} & 8.01 \small{$\pm 0.67$} & 1.93 \small{$\pm 0.11$} & 10.31 \small{$\pm 0.19$} & -- \\
+ UMA relaxation (10) & 10 & 2462 & 8 & 93.65 \small{$\pm 0.40$} & 4.41 \small{$\pm 0.46$} & 1.95 \small{$\pm 0.10$} & 10.39 \small{$\pm 0.17$} & -- \\
+ UMA relaxation (50) & 50 & 10283 & 8 & 93.35 \small{$\pm 0.41$} & 4.63 \small{$\pm 0.40$} & 1.97 \small{$\pm 0.10$} & 10.44 \small{$\pm 0.18$} & -- \\
+ PCFM projection & 0 & 396 & 8 & 93.11 \small{$\pm 0.42$} & 9.29 \small{$\pm 0.76$} & 1.83 \small{$\pm 0.10$} & 10.20 \small{$\pm 0.14$} & -- \\
+ parity, MMFF, declash & 0 & 930 & 8 & 93.43 \small{$\pm 0.33$} & 2.80 \small{$\pm 0.47$} & 1.93 \small{$\pm 0.10$} & 10.78 \small{$\pm 0.16$} & \best{95.1} \\

\bottomrule
\end{tabular}%
}
\end{table}

\begin{figure}[t]
\centering
\includegraphics[width=0.9\textwidth]{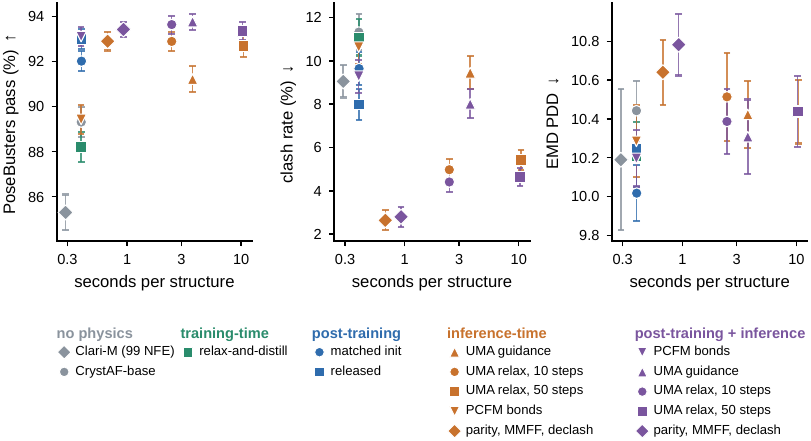}
\caption{Quality versus inference cost for Table~\ref{tab:main}. Post-training improves CrystAF-base with essentially no added inference cost, while inference-time corrections trade cost for better validity and fewer clashes; combining both gives the lowest clash rates. Colour denotes route, marker shape method, and error bars bootstrap SE.}
\label{fig:cost}
\end{figure}

Table~\ref{tab:main} and Figure~\ref{fig:cost} put the three routes side by side, and they fix different things.

\textbf{Post-training improves the whole packing, and sampling stays as fast.} From the same starting model, CrystAF-UMA raises PB from $89.3\%$ to $92.0\%$ and lowers the clash rate from $11.3\%$ to $9.6\%$. It is also the only route that moves the cell volume ($1.95\to1.69$) and the distance distribution ($10.44\to10.02$) toward the experimental crystal. The architecture and the 32 network evaluations are unchanged, so each structure costs the same as before. The price, about 5 GPU-hours, is paid once.

\textbf{Relaxation is better at local repair, but it pays on every structure.} Ten fixed-cell UMA steps on CrystAF-base give $92.9\%$ PB and cut clashes to $5.0\%$. Because the cell is fixed, volume error stays at $1.95$ and PDD slightly worsens to $10.51$. The ten UMA calls make sampling $6.2\times$ slower ($2.5$\,s vs.\ $0.40$\,s). Fifty steps do no better ($92.7\%$ PB, $5.4\%$ clash) at $26\times$ cost, showing that the gain saturates within ten steps. Force guidance is harder to tune: the gentle setting gives $91.2\%$ PB and $9.5\%$ clash with little change in cell or PDD, yet costs $3.8$\,s per structure because each guided jump requires a potential call. Strong guidance trades pass rate for fewer clashes ($90.4\%$ PB, $6.8\%$ clash) and worsens PDD to $10.53$. Projecting bond lengths without chirality changes almost nothing on either model ($89.4\%$ on CrystAF-base, $93.1\%$ on CrystAF-UMA), indicating that the 15-point loss of full PCFM comes from its chirality term.

\textbf{Training on relaxed targets does not transfer the repair.} Relax-and-distill sees the same potential and the same families, yet it reaches only $88.2\%$ PB with a small packing gain (Vol.Err $1.74$, PDD $10.21$). Regressing each sample onto its relaxed copy asks the model to reproduce 25 descent steps it never takes at sampling time. A ranking reward asks for much less: prefer some of your own samples over others.

\textbf{The routes stack.} Inference-time correction does not care where a sample came from; it fixes contacts and stereocentres and leaves the rest alone. On CrystAF-UMA, the classical chain lowers the clash rate to $2.8\%$ and raises stereochemical agreement from $48\%$ to $95\%$, with PB at $93.4\%$. Ten relaxation steps give $93.6\%$ PB and $4.4\%$ clash, better than either route alone on both counts, and fifty steps add nothing ($93.4\%$, $4.6\%$). Gentle guidance adds $0.8$ points of PB ($93.8\%$). Meanwhile, CrystAF-base with the same classical chain reaches $92.9\%$ PB and a PDD of $10.64$. Local correction does not recover what post-training adds to the packing, and on top of post-training it is cheap.

\subsection{What the model learns from the signal}
\label{sec:what-learned}

\begin{table}[t]
\caption{Ablations of the physical signal and update rule. The first two blocks start from CrystAF-base and are read at epoch 4. The last block compares the UMA reward with a PoseBusters-ranked reward from a shared warm-started checkpoint (seed 929); second seeds are in Table~\ref{tab:seeds}. Results use 200 families $\times$ 20 samples at 32 NFE. Epoch-6 results are in Table~\ref{tab:ablation-ep6}.}
\label{tab:ablation}
\centering
\small
\setlength{\tabcolsep}{2.5pt}
\resizebox{\textwidth}{!}{%
\begin{tabular}{lccccc}
\toprule
& & \multicolumn{4}{c}{Quality} \\
\cmidrule(lr){3-6}
Variant & Epoch & PB \% $\uparrow$ & Clash \% $\downarrow$ & Vol.Err $\downarrow$ & PDD $\downarrow$ \\
\midrule

\rowcolor{gray!10}
\multicolumn{6}{c}{\textit{Signal ablations --- matched CrystAF-base initialization}} \\
CrystAF-base (no post-training) & --- & 89.32 \small{$\pm$0.67} & 11.33 \small{$\pm$0.84} & 1.95 \small{$\pm$0.09} & 10.44 \small{$\pm$0.16} \\
+ $E$, $F$, $\sigma$, eligibility & 4 & 91.21 \small{$\pm$0.46} & 9.42 \small{$\pm$0.75} & 1.66 \small{$\pm$0.09} & 9.99 \small{$\pm$0.15} \\
+ $E$, eligibility & 4 & 89.01 \small{$\pm$0.52} & 9.39 \small{$\pm$0.75} & 1.55 \small{$\pm$0.08} & 9.94 \small{$\pm$0.11} \\
+ Eligibility, volume & 4 & 91.52 \small{$\pm$0.46} & 9.82 \small{$\pm$0.77} & 1.62 \small{$\pm$0.09} & 9.98 \small{$\pm$0.12} \\
+ Eligibility & 4 & 92.10 \small{$\pm$0.47} & 10.35 \small{$\pm$0.82} & 1.83 \small{$\pm$0.10} & 10.13 \small{$\pm$0.18} \\
+ $E$, $F$, $\sigma$, eligibility, volume (CrystAF-UMA) & 4 & 92.02 \small{$\pm$0.44} & 9.64 \small{$\pm$0.75} & 1.69 \small{$\pm$0.09} & 10.02 \small{$\pm$0.14} \\

\midrule
\rowcolor{gray!10}
\multicolumn{6}{c}{\textit{Update and training baselines --- matched CrystAF-base initialization}} \\
DiffusionNFT on $U$, full signal & 4 & 92.66 \small{$\pm$0.44} & 10.23 \small{$\pm$0.79} & 2.51 \small{$\pm$0.11} & 10.71 \small{$\pm$0.17} \\
Relax-and-distill & 4 & 88.20 \small{$\pm$0.68} & 11.08 \small{$\pm$0.83} & 1.74 \small{$\pm$0.09} & 10.21 \small{$\pm$0.18} \\

\midrule
\rowcolor{gray!10}
\multicolumn{6}{c}{\textit{Reward comparison --- shared warm-started initialization, seed 929}} \\
CrystAF-UMA & 6 & 93.01 \small{$\pm$0.43} & 7.98 \small{$\pm$0.70} & 1.88 \small{$\pm$0.10} & 10.25 \small{$\pm$0.20} \\
PB-rank reward & 12 & 93.21 \small{$\pm$0.42} & 7.86 \small{$\pm$0.67} & 1.61 \small{$\pm$0.11} & 10.07 \small{$\pm$0.23} \\

\bottomrule
\end{tabular}%
}
\end{table}

Table~\ref{tab:ablation} changes one thing at a time. Every arm starts from CrystAF-base, follows the same recipe, and is read out at epoch 4.

\textbf{The full signal improves all four metrics.} Post-training with the full signal raises PB from $89.3\%$ to $92.0\%$, lowers the clash rate from $11.3\%$ to $9.6\%$, and reduces the volume error from $1.95$ to $1.69$ and PDD from $10.44$ to $10.02$. The PDD change matters most. PDD compares the whole distance distribution with the experimental crystal, it is the metric furthest from anything in the signal, and it moves by about 2.5 SE. The released CrystAF-UMA model, which starts from a warm-started checkpoint, reaches a higher pass rate ($93.0\%$ and $93.5\%$ over two seeds, Table~\ref{tab:seeds}) but a smaller PDD gain.

\textbf{The gate fixes validity; the potential improves packing.} The gate-only arm keeps only the eligibility check: implausible candidates receive the lowest advantage, while all plausible ones are treated alike. This already gives the full PoseBusters gain ($92.1\%$ versus $92.0\%$ for the full signal). The potential adds a ranking among plausible candidates, improving packing. With energy, forces, and stress but no volume term, clashes fall from $10.3\%$ to $9.4\%$, volume error from $1.83$ to $1.67$, and PDD from $10.13$ to $10.00$. Each change is about one SE and points in the same direction. Energy alone drifts fastest (KL $0.7$ by epoch 4, Table~\ref{tab:train-signals}) and leaves PB flat, but gives the best volume error and PDD ($1.55$, $9.94$). A volume term without the potential also recovers most of the packing gain ($1.62$, $9.98$), because it targets volume directly.

\textbf{The update matters as much as the signal.} DiffusionNFT applied to $U$ with the full signal reaches a similar pass rate ($92.7\%$), but the rest of the picture reverses. Volume error rises to $2.51$ and PDD to $10.71$, both worse than CrystAF-base. This is exactly the narrow kind of gain we wanted to tell apart from a physical one: the pass rate goes up while the packing moves away from experiment. What separates the two runs is the flow-map conversion of Equation~\ref{eq:identity}. The conversion also makes training less stable (Table~\ref{tab:train-signals}), which is why every arm is read out before the drift sets in.

\textbf{What the potential itself sees.} If post-training simply moved samples downhill on UMA, their energy would drop. It rises instead. Every arm trained with our update ends $0.8$--$1.6$\,eV per molecule higher in single-point energy than CrystAF-base, with more stress, even when rewarded on energy. Its cells nevertheless end closer to the experimental volume (Appendix~\ref{sec:uma-diag}). Two effects explain this. UMA's equilibrium cell is larger than the experimental one, so a cell moving toward experiment looks compressed in an unrelaxed single point. A group-relative advantage also has no handle on absolute energy, which drifts even for the gate-only arm. Relaxation, by contrast, lowers energy by $3.6$\,eV per molecule while leaving the cell unchanged. Post-training thus buys agreement with experiment, which CSP needs, rather than progress on UMA's own objective.

\textbf{A reward built on the metric.} The last block starts both rewards from the same warm-started checkpoint. Ranking candidates by PoseBusters itself reaches $93.2\%$ after 12 epochs; the UMA reward reaches $93.0\%$ after 6, with similar clash, volume error, and PDD. A second seed gives the same picture ($92.6\%$ against $93.5\%$, Table~\ref{tab:seeds}). Optimizing the proxy directly buys nothing on it that physical feedback does not already give.

\subsection{Some physics should stay at inference time}
\label{sec:stereo}

Stereochemistry shows where learning stops. Clari's backbone is conditioned on the molecular graph, and the graph is the same for both mirror images, so no reward can tell the model which one was requested. Across 20 training configurations, agreement with the requested R/S labels stayed at $49$--$53\%$. At inference time the problem is nearly free to fix. Per-molecule parity inversion reaches $85\%$ with PoseBusters unchanged, and adding restrained MMFF and a declash step brings it to $95\%$ (Figure~\ref{fig:stereo}). A Gauss--Newton projection reaches $99.97\%$, at a cost of 15 PoseBusters points. Cell calibration is a similar case: a single $3\%$ volume offset, fitted on the training split, is easier to apply at the end than to learn.

\subsection{Crystal recovery}
\label{sec:recovery}

PoseBusters and PDD measure whether a sample looks like a crystal. CSP needs something harder: the right crystal among the proposals. We follow Clari's structure-solution benchmark. For each held-out target we draw $n_s$ candidates, keep the $k$ with the lowest UMA energy, and count the target as solved if any of them matches the experimental structure under COMPACK \citep{motherwell2025compack}: at least 8 of 15 molecules within $2$\,\AA{} RMSD, with no collision. Nothing in the training signal refers to these experimental structures. Table~\ref{tab:solved} uses 150 candidates per target and keeps the 30 lowest in energy, a budget close to how a CSP campaign would screen; it covers the rigid, flexible, and CSP blind-test subsets of OXtal and the 779-target teaching set.

\begin{table}[t]
\caption{Structure recovery with $n_s=150$ candidates and $k=30$ lowest by UMA energy ($\mathrm{Sol}@k$; mean $\pm$ SE). ``+ chain'' applies parity inversion, restrained MMFF, rigid declashing, and cell calibration. Clari-M uses 39 NFE; CrystAF uses 32 NFE. CSP subsets contain only 5--8 targets, so target-level uncertainty exceeds the reported SE. \textbf{Bold}: best per column.}
\label{tab:solved}
\centering
\small
\setlength{\tabcolsep}{3.5pt}
\resizebox{\textwidth}{!}{%
\begin{tabular}{lcccccc}
\toprule
& Rigid (50) & Flexible (50) & CSP5 (6) & CSP6 (5) & CSP7 (8) & Teaching (779) \\
\midrule
Clari-M & $0.685{\pm}0.033$ & $0.236{\pm}0.033$ & $0.331{\pm}0.021$ & $0.327{\pm}0.096$ & $0.234{\pm}0.043$ & $0.441{\pm}0.007$ \\
\midrule
CrystAF-base & $0.693{\pm}0.027$ & $0.276{\pm}0.030$ & $0.490{\pm}0.040$ & $0.200{\pm}0.000$ & $\mathbf{0.267}{\pm}0.094$ & $0.456{\pm}0.007$ \\
CrystAF-base + chain & $\mathbf{0.732}{\pm}0.026$ & $0.308{\pm}0.032$ & $0.431{\pm}0.087$ & $0.455{\pm}0.136$ & $0.240{\pm}0.035$ & $\mathbf{0.486}{\pm}0.008$ \\
\midrule
CrystAF-UMA & $0.705{\pm}0.027$ & $0.219{\pm}0.024$ & $0.479{\pm}0.055$ & $0.327{\pm}0.096$ & $0.246{\pm}0.021$ & $0.451{\pm}0.007$ \\
CrystAF-UMA + chain & $0.708{\pm}0.028$ & $\mathbf{0.310}{\pm}0.027$ & $\mathbf{0.583}{\pm}0.100$ & $\mathbf{0.519}{\pm}0.105$ & $0.248{\pm}0.014$ & $0.468{\pm}0.007$ \\
\bottomrule
\end{tabular}}
\end{table}

Two patterns stand out. First, the inference-time chain improves recovery on both models. On CrystAF-base it raises the teaching score from $0.456$ to $0.486$, rigid from $0.693$ to $0.732$, and flexible from $0.276$ to $0.308$. On CrystAF-UMA it improves every column, lifting flexible from $0.219$ to $0.310$ and the CSP5 and CSP6 blind tests from $0.479$ to $0.583$ and from $0.327$ to $0.519$. Part of the reason is mechanical: a solved target needs a collision-free match, and declashing turns near-misses into hits. The chain changes which candidates count, whichever model produced them. Second, CrystAF-base and CrystAF-UMA trade wins. CrystAF-UMA is ahead on rigid targets ($0.705$ vs.\ $0.693$) and on CSP6, behind on flexible ones ($0.219$ vs.\ $0.276$) and on CSP7, and level on the teaching set ($0.451$ vs.\ $0.456$). With 400 candidates the wins move around but the mix remains: CrystAF-UMA leads on flexible targets and CSP5 and trails on CSP7 and the teaching set (Table~\ref{tab:solved-ns400}). Post-training makes each sample more plausible; it does not, by itself, change which experimental packings the model finds. Every CrystAF row still matches or beats Clari-M on the teaching set.

\section{Discussion and Conclusion}
\label{sec:discussion}

\textbf{When to learn physics and when to apply it.}
Post-training costs about 5 GPU-hours once, whereas ten UMA relaxation steps add about $2.1$\,s per structure. The break-even point is roughly $8{,}700$ structures, or 22 molecules at 400 candidates each. Learning physics is therefore cheaper for repeated use, while inference-time correction is preferable when the physical model changes often or the generator cannot represent the constraint, as with stereochemistry.

\textbf{Physical improvement or metric optimization?}
A higher pass rate alone says little: DiffusionNFT and a PoseBusters-ranked reward both reach it, while DiffusionNFT moves the packing away from experiment. The stronger evidence is that pass rate, cell volume, and PDD, none seen by the reward, improve together only with the flow-map-aware update. This agreement is with experiment rather than the potential, whose sample energies increase. Physical feedback therefore mainly identifies broken samples and ranks the plausible ones to improve packing. The same post-training also transfers: on public Clari-M, with no PoseBusters step in its history, PB rises from $88.4\%$ to $91.6\%$; on rigid-body MolCrystalFlow, volume error falls from $3.88\%$ to $3.16\%$, where NFT also outperforms Flow-GRPO (Appendix~\ref{sec:generality-details}).

\textbf{Local physics is not global search.}
Post-training changes what a sample looks like, not which packings the model proposes. Against CrystAF-base, it wins on some recovery subsets and loses on others (Table~\ref{tab:solved}), while the inference-time chain helps both models. A rank-based objective that reweights the model's own samples cannot create packings the model never samples. Finding the right packing more often will take something else, such as stronger conditioning, rewards that favor diversity, or higher-fidelity energies for ranking.

% \section{Conclusion}

Holding the generator and physical signal fixed, we find that where physics enters determines both what improves and what it costs. Post-training learns reusable physical preferences into the generator, improving validity and packing with no added sampling cost, whereas relaxation and other inference-time corrections repair residual local defects but pay per structure. Training on relaxed targets provides little benefit. Much of the validity gain comes from identifying broken samples, while the update rule determines whether the improvement extends beyond the optimized signal. The routes are complementary: post-training improves the generated distribution, and inference-time correction further reduces clashes, restores stereochemistry, and can convert near-misses into recovered experimental structures. The same post-training recipe also transfers across all-atom and rigid-body generators. Together, these results argue for learning broadly reusable physics into the model and reserving inference-time correction for constraints and repairs that are better enforced than learned.

\subsection*{AI Use Statement}
Generative AI tools were used to assist with manuscript editing, language refinement, LaTeX formatting, and mathematical consistency checking. They were not used as a substitute for experimental validation or independent verification of the reported results. The authors reviewed and verified all AI-assisted text, derivations, implementation details, references, and experimental claims, and take full responsibility for the final content of the paper.

\subsection*{Ethics statement}
This work uses crystallographic data from the Cambridge Structural Database (CSD) under the applicable CCDC license; the licensed data are not redistributed with the released artifacts. The generators and their UMA-based scores inherit limitations of their training data and pretrained components, and the reported computational metrics do not constitute experimental validation.

\section*{Reproducibility statement}
The evaluation protocol, grid exponents, read-out epoch, and cost measurement are specified in Section~\ref{sec:results} and Appendix~\ref{sec:cost}. Every post-training arm, the training-time baseline, and every training-free method has a configuration file or an environment-variable recipe in the released code, together with the figure data. We release CrystAF-base, CrystAF-UMA, and the distillation teacher. The evaluation set is derived from the Cambridge Structural Database and cannot be redistributed; the repository documents how to build it from a CSD licence. COMPACK requires a CSD-Materials licence.  The code is available on \href{https://github.com/HaCTang/CrystAF}{https://github.com/HaCTang/CrystAF}, and the model weights are available on \href{https://huggingface.co/Haocheng1/CrystAF}{https://huggingface.co/Haocheng1/CrystAF}.

\bibliography{iclr2027_conference}
\bibliographystyle{iclr2027_conference}

\clearpage
\appendix
% Appendix floats are numbered S1, S2, ... so they can never be confused with
% the main-body tables and figures when they are cited from the text.
\setcounter{table}{0}
\setcounter{figure}{0}
\renewcommand{\thetable}{S\arabic{table}}
\renewcommand{\thefigure}{S\arabic{figure}}

\section{Appendix}

\subsection*{A.\quad Protocol and cost}

\subsection{Data split and validation protocol}
\label{sec:data-splits}

We follow Clari's original data split without modification. Clari groups CSD records by the first six refcode characters, keeping polymorphs within one split; it assigns the published OXtal and teaching families to test, removes training entries with an exact asymmetric-unit component-SMILES match to a retained test component of more than seven heavy atoms, then samples 1,000 validation families from the remaining families with \texttt{random.Random(42)}. Training and volume calibration use \texttt{train.pt}; Table~\ref{tab:solved} uses the Clari test families.

Following Clari's fixed validation protocol, Table~\ref{tab:main} uses the first 200 deterministic representatives in \texttt{val.pt}. We use this same subset for matched model selection and reporting, including the reported epoch, budget-specific $\rho$, and correction recipe.

\subsection{Protocol effects: the same checkpoints on both validation sets}

Table~\ref{tab:protocol} is the reason every number in the body is quoted on one protocol. The 200-family subset used throughout this paper is harder than the full 1000-family validation set --- the same Clari-M checkpoint loses $3.1$ points of PoseBusters pass rate and $0.37$ of EMD PDD on it --- so a CrystAF row measured at 200 families cannot be compared against a published 1000-family baseline without understating CrystAF by roughly that much. Our 1000-family reproduction of both backbones is within $1.6$ PB of the published figures, which is what licenses treating the 200-family gap as a property of the subset rather than of our evaluation code.

\begin{table}[h]
\caption{Clari-M and Clari-L results on the 1000-family and 200-family validation protocols. Both use 50 Heun steps, 20 samples per family, and identical metric code; ``published'' is the original paper's Table 1.}
\label{tab:protocol}
\centering
\small
\begin{tabular}{llrrrr}
\toprule
& protocol & PB \% $\uparrow$ & clash \% $\downarrow$ & Vol.Err $\downarrow$ & EMD PDD $\downarrow$ \\
\midrule
\multirow{3}{*}{Clari-M}
 & published, 1000 fam & 87.34 & 9.56 & 1.59 & 9.56 \\
 & ours, 1000 fam & 88.43 \small{$\pm$0.33} & 8.57 \small{$\pm$0.34} & 1.69 \small{$\pm$0.04} & 9.82 \small{$\pm$0.08} \\
 & \textbf{ours, 200 fam} & 85.30 \small{$\pm$0.80} & 9.05 \small{$\pm$0.74} & 1.83 \small{$\pm$0.13} & 10.19 \small{$\pm$0.36} \\
\midrule
\multirow{3}{*}{Clari-L}
 & published, 1000 fam & 85.89 & 7.69 & 1.50 & 9.28 \\
 & ours, 1000 fam & 86.88 \small{$\pm$0.39} & 6.92 \small{$\pm$0.30} & 1.53 \small{$\pm$0.04} & 9.43 \small{$\pm$0.04} \\
 & \textbf{ours, 200 fam} & 85.00 \small{$\pm$0.86} & 7.75 \small{$\pm$0.68} & 1.60 \small{$\pm$0.09} & 9.77 \small{$\pm$0.14} \\
\bottomrule
\end{tabular}
\end{table}

\subsection{How cost is measured}
\label{sec:cost}

Inference cost is measured by generating the first 4 validation families $\times$ 20 samples with each method on one L40S GPU (20 structures per sampler call, bfloat16 network, float32 UMA). The first call, which includes CUDA warm-up and loading UMA, is dropped, so every method is timed on the same three families. The other GPUs of the node were running evaluations at the time, which affects CPU-bound steps such as MMFF. Absolute times depend on the families drawn: one of the three has a large molecule and takes most of the time. The ratios between methods are the meaningful quantity. We report wall-clock milliseconds per generated structure, synchronized around the sampler call, so metric computation is excluded while every sampling-time correction is included. MMFF, parity inversion, and the declash step run on the CPU inside the same call and are counted. UMA calls are counted per structure: force guidance makes one call per guided jump, and relaxation one call per step. Training cost is the L40S GPU-hours spent on top of CrystAF-base: four epochs on four GPUs take 71 minutes for CrystAF (4.7 GPU-hours) and 90 minutes for relax-and-distill (6.0 GPU-hours). The released CrystAF-UMA model took six epochs (about 8 GPU-hours) on top of its warm start. Distilling CrystAF-base itself is shared by every row and is not counted.

\paragraph{Clari-M against CrystAF.} Per sampler call of 20 structures, CrystAF-base takes $2.1$, $18.1$, and $3.6$\,s on the three timed families, and Clari-M (99 NFE) takes $5.1$, $9.8$, and $2.4$\,s. On the two smaller molecules CrystAF is faster, as its lower NFE predicts. On the largest it is slower, and we have not identified why. Its average over the three families is therefore higher than Clari-M's (396 vs.\ 289\,ms). We make no wall-clock claim for CrystAF over Clari-M. All cost ratios in the main text compare methods on the same CrystAF weights.

\subsection{What the signal shares with the metrics}
\label{sec:overlap}

\begin{table}[H]
\caption{Overlap between the physical signal and the evaluation metrics. \checkmark: the metric is directly part of the signal; $\sim$: related but different quantity; ---: not seen.}
\label{tab:overlap}
\centering
\small
\setlength{\tabcolsep}{4pt}
\begin{tabular}{lccccc}\toprule
signal term & PB pass & clash (cov.\ radii) & Vol.Err & EMD PDD & COMPACK \\
\midrule
UMA energy, forces, stress & --- & $\sim$ & $\sim$ & --- & --- \\
eligibility ($d_{\min}\ge 0.8$\,\AA, density) & --- & $\sim$ & $\sim$ & --- & --- \\
cell-volume term (optional) & --- & --- & \checkmark & $\sim$ & --- \\
\bottomrule
\end{tabular}
\end{table}

\subsection*{B.\quad The generator}

\subsection{Diagnostics: jump length, sampling budget, stereochemistry}
\begin{figure}[H]
\centering
\begin{subfigure}[b]{0.32\textwidth}
\includegraphics[width=\textwidth]{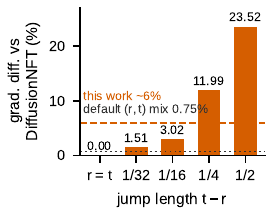}
\caption{}
\label{fig:dose}
\end{subfigure}
\hfill
\begin{subfigure}[b]{0.32\textwidth}
\includegraphics[width=\textwidth]{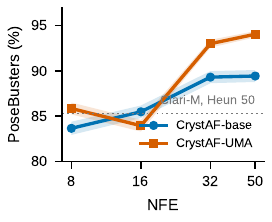}
\caption{}
\label{fig:nfe}
\end{subfigure}
\hfill
\begin{subfigure}[b]{0.32\textwidth}
\includegraphics[width=\textwidth]{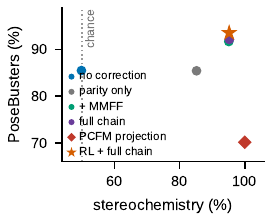}
\caption{}
\label{fig:stereo}
\end{subfigure}
\caption{(a) How much of the post-training gradient comes from the flow-map term of Equation~\ref{eq:identity}, as a function of jump length (Equation~\ref{eq:gradient-difference}). (b) PoseBusters pass rate across sampling budgets for one set of weights, before and after post-training. (c) Stereochemistry and validity for inference-time recipes on the same weights.}
\end{figure}

\subsection{Distillation schedule and compute accounting}
\label{sec:si_distillation}

\begin{figure}[H]
\centering
\begin{subfigure}[b]{0.505\textwidth}
\includegraphics[width=\textwidth]{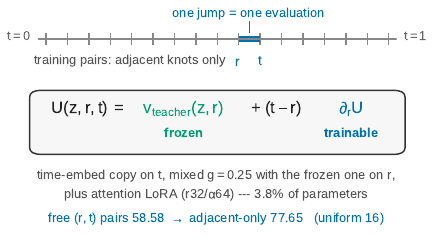}
\caption{Distillation on adjacent grid knots.}
\label{fig:si_distillation_schedule}
\end{subfigure}
\hfill
\begin{subfigure}[b]{0.475\textwidth}
\includegraphics[width=\textwidth]{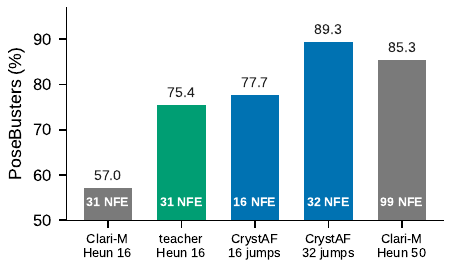}
\caption{Compute-matched sampling budgets.}
\label{fig:si_distillation_budget}
\end{subfigure}
\caption{CrystAF distillation and compute accounting. (a) The student learns a jump-scaled residual correction relative to the frozen teacher, with adjacent grid knots matching deployed transitions. (b) Network-evaluation counts and PoseBusters validity for flow-map jumps and Heun sampling.}
\label{fig:si_distillation}
\end{figure}

The teacher is Clari-M after NFT post-training on a PoseBusters-ranked reward at 16 Heun steps; every CrystAF row inherits it, so the main text compares only against CrystAF-base, never against Clari-M (a comparison with no PoseBusters-derived step is the Clari-M row of Table~\ref{tab:generality}). Figure~\ref{fig:si_distillation} shows the two implementation details omitted from the main text. The dual-time student is first exposed to unrestricted $(r,t)$ pairs for 2k updates and is then trained for 6k updates on adjacent knots of a 16-step grid. The adjacent-jump phase aligns the training intervals with the transitions used at inference and raises uniform 16-step validity from $58.6\%$ to $77.7\%$. Throughout distillation, the frozen teacher supplies the reference velocity, while the trainable target-time pathway and attention LoRA learn the jump-scaled residual correction.

The two samplers use different accounting. One CrystAF jump requires one network evaluation, so an $N$-jump trajectory costs $N$ evaluations. A $T$-step Heun trajectory costs $2T-1$ evaluations because it uses a predictor and a corrector, with the final evaluation shared. We therefore compare methods by NFE rather than by the nominal number of integration steps. This convention is also used in Tables~\ref{tab:main} and~\ref{tab:grid}.

The adapter weights are shared across $N\in\{8,16,32,50\}$, but the sampling grid remains budget dependent through $t_i=(i/N)^\rho$. The selected exponents are $\rho=0.30$, $0.75$, $1$, and $1$ for 8, 16, 32, and 50 NFE, respectively. This is a sampling-time choice rather than a change of model parameters: at 8 NFE alone, using $\rho=0.30$ instead of $0.75$ increases validity from $48.3\%$ to $83.7\%$. The result explains why a single any-step adapter still requires a grid matched to its deployment budget.

\begin{figure}[t]
\centering
\includegraphics[width=\textwidth]{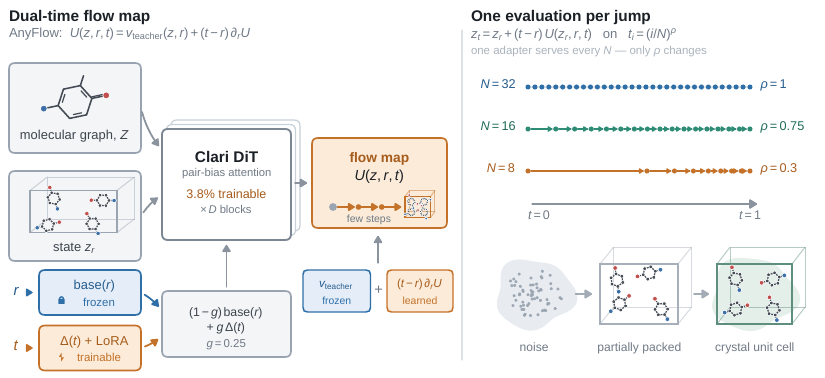}
\caption{Dual-time flow-map architecture and sampling grid used by CrystAF. \textbf{Left}: the frozen Clari time embedder reads state time $r$; a trainable copy and attention LoRA read target time $t$. The map combines the teacher velocity with a learned jump-scaled correction. \textbf{Right}: each jump costs one network evaluation, with $t_i=(i/N)^\rho$ for different budgets.}
\label{fig:arch}
\end{figure}

\subsection{Sampling budgets with one adapter}

\begin{table}[t]
\caption{Results for four sampling budgets with shared adapters. Weights are identical within each block; only $t_i=(i/N)^\rho$ changes. Uniform 16-step sampling scores $77.65$ and the tuned grid scores $85.49$; \textbf{bold} is best per column.}
\label{tab:grid}
\centering
\small
\setlength{\tabcolsep}{5pt}
\begin{tabular}{llcccc}
\toprule
& NFE ($\rho$) & PB \% $\uparrow$ & clash \% $\downarrow$ & Vol.Err $\downarrow$ & EMD PDD $\downarrow$ \\
\midrule
\multirow{4}{*}{CrystAF-base}
 & 8 (0.30) & 83.67 & 30.73 & 2.46 & 11.54 \\
 & 16 (0.75) & 85.49 & 13.85 & 2.09 & 10.92 \\
 & 32 (1) & 89.32 & 11.33 & 1.95 & 10.44 \\
 & 50 (1) & 89.44 & 10.88 & 1.86 & 10.47 \\
\midrule
\multirow{4}{*}{\shortstack[l]{CrystAF-UMA}}
 & 8 (0.30) & 85.86 & 35.07 & 3.66 & 13.26 \\
 & 16 (0.75) & 83.96 & 11.95 & 2.17 & 10.71 \\
 & 32 (1) & 93.01 & 7.98 & 1.88 & 10.25 \\
 & 50 (1) & \best{94.05} & 8.13 & 1.84 & 10.24 \\
\midrule
\multirow{4}{*}{\shortstack[l]{PB-rank reward}}
 & 8 (0.30) & 86.35 & 27.17 & 2.63 & 11.69 \\
 & 16 (0.75) & 85.71 & 11.29 & 1.85 & 10.45 \\
 & 32 (1) & 93.21 & 7.86 & \best{1.61} & \best{10.07} \\
 & 50 (1) & 93.84 & \best{7.37} & 1.70 & 10.22 \\
\bottomrule
\end{tabular}
\end{table}

Figure~\ref{fig:nfe} and Table~\ref{tab:grid} show that the same parameters can be used from 8 to 50 evaluations once the grid is adapted. The distilled model rises from $83.67\%$ validity at 8 NFE to $89.32\%$ at 32 NFE, while CrystAF-UMA reaches $93.01\%$ at 32 and $94.05\%$ at 50. Post-training gains are concentrated at the 32--50 NFE operating points; at 16 NFE the physical-reward checkpoint is roughly level with the distilled student on validity while improving clash and PDD. In other words, parameter sharing across budgets does not imply budget-invariant alignment: the flow map is any-step, but the RL distribution should still be chosen for the deployment regime.

\subsection*{C.\quad Post-training details}

\subsection{Complete physical-reward and NFT objective}
\label{sec:objective-details}

This section makes explicit the objective summarized in \S\ref{sec:postrain}. Training first samples a molecular family $g$ uniformly from the training-family set and draws a group of $K$ terminal candidates $x_{1,i}\sim\pi_{\mathrm{old}}(\cdot\mid g)$, $i=1,\ldots,K$, where $\pi_{\mathrm{old}}$ is the EMA data-collection policy.  A separate frozen anchor is used only by the KL surrogate below.  For the physical arm, we orient every raw channel so that larger is better:
\begin{equation}
\begin{aligned}
q^E_{gi} &= -E_{gi}^{\mathrm{mol}},\qquad
q^{\bar F}_{gi} = -\bar F_{gi},\qquad
q^{F_{\max}}_{gi} = -\log\!\left(1+F^{\max}_{gi}\right), \\
q^\sigma_{gi} &= -\lVert \sigma_{gi}\rVert_F,\qquad
q^V_{gi} = -\left|V_{gi}/V_g^\star-1\right|.
\end{aligned}
\label{eq:reward-channels}
\end{equation}
where $E_{gi}^{\mathrm{mol}}$ is the UMA energy per molecule and $V_g^\star$ is the family target volume.  Here $\bar F_{gi}$ and $F^{\max}_{gi}$ are the mean and maximum atomic-force norms. Thus lower energy, force, stress, and volume error all increase desirability.  The scorer's geometry checks and finite energy requirement define the eligibility indicator
\begin{equation}
m_{gi}=\mathbb I[\mathrm{valid}_{gi}]
\mathbb I\!\left[E^{\mathrm{mol}}_{gi}\ \text{is finite and not failed}\right].
\label{eq:eligibility}
\end{equation}
The scorer sets $\mathrm{valid}_{gi}$ from the minimum-distance threshold and fixed dataset-level density bounds; these bounds are not estimated from the evaluation set.

The order of operations is important. For $c\in\{E,\bar F,F_{\max},\sigma\}$, entries rejected by $m_{gi}$ are replaced by that channel's mean over valid candidates before standardisation. This prevents the failed-score sentinel from determining the group mean and variance. The volume channel is defined for every candidate and is standardised directly. Let $\widetilde q^c_{gi}$ denote the resulting value (with $\widetilde q^V_{gi}=q^V_{gi}$) and define, for $c\in\{E,\bar F,F_{\max},\sigma,V\}$,
\begin{equation}
z^c_{gi}=\frac{\widetilde q^c_{gi}-\mu^c_g}
{s^c_g+\epsilon_z},\qquad
\mu^c_g=\frac1K\sum_{j=1}^K\widetilde q^c_{gj},\quad
(s^c_g)^2=\frac1K\sum_{j=1}^K(\widetilde q^c_{gj}-\mu^c_g)^2.
\label{eq:within-family-standardization}
\end{equation}
In the reported UMA runs, maximum force and stress are enabled as channels; their non-zero configuration values act as switches after standardisation rather than multiplicative weights. With $c_{gi}\in\{0,1\}$ the clash indicator and $h_{gi}\in[-1,0]$ the optional clearance term, the pre-centering physical score is
\begin{equation}
\begin{aligned}
\widetilde S_{gi}
&= \lambda_{\mathrm{ef}} z^E_{gi}
+(1-\lambda_{\mathrm{ef}}) z^{\bar F}_{gi}
+\mathbb I[w_{F_{\max}}\ne 0] z^{F_{\max}}_{gi}
+\mathbb I[w_\sigma\ne 0] z^\sigma_{gi} \\
&\quad
+w_{\mathrm{vol}} z^V_{gi}
-w_{\mathrm{clash}} c_{gi}
+w_{\mathrm{margin}} h_{gi},
\qquad
S_{gi}
=\widetilde S_{gi}
-\operatorname*{median}_{j:m_{gj}=1}\widetilde S_{gj}.
\end{aligned}
\label{eq:physical-score}
\end{equation}
Groups with no valid candidate are skipped. For the reported volume-guided UMA runs, we then clip and apply the non-positive rejection gate,
\begin{equation}
A_{gi}=
\begin{cases}
\operatorname{clip}(S_{gi},-1,1), 
    & m_{gi}=1,\\
\min\!\left\{\operatorname{clip}(S_{gi},-1,1),0\right\}, 
    & m_{gi}=0,
\end{cases}
\qquad
p_{gi}=\frac{A_{gi}+1}{2}\in[0,1].
\label{eq:rejection-consistent-advantage}
\end{equation}
Consequently, rejection is decided from geometry and scoreability, rejected entries in UMA channels are repaired before group normalisation, and the final gate prevents an invalid candidate with a favourable volume or energy score from acquiring positive advantage. The PoseBusters arm replaces Equation~\ref{eq:physical-score} by its within-family PB rank (top and bottom quartiles); it retains the same clash veto and cell term.

For each candidate we draw $\epsilon_i\sim\mathcal N(0,I)$ and use the linear forward path
\begin{equation}
x_{s,i}=(1-s)\epsilon_i+s x_{1,i},\qquad
w_i=\frac{d x_{s,i}}{ds}=x_{1,i}-\epsilon_i.
\label{eq:forward-path}
\end{equation}
For the reported jump-aware runs, let $\tau_j=(j/N)^\rho$ be the configured power grid. We draw $j$ uniformly from $\{0,\ldots,N-1\}$ and set $(r,t)=(\tau_j,\tau_{j+1})$, so the training distribution contains no zero-length intervals. Let $\widehat D_rU_{\mathrm{old}}$ be the shared derivative estimator in Equation~\ref{eq:central-total-derivative}. The trainable, EMA old-policy, and frozen-anchor induced velocities are
\begin{align}
V_\theta&=U_\theta(x_{r,i},r,t)
 -(t-r)\operatorname{sg}\!\left[\widehat D_rU_{\mathrm{old}}\right],\\
V_{\mathrm{old}}&=U_{\mathrm{old}}(x_{r,i},r,t)
 -(t-r)\operatorname{sg}\!\left[\widehat D_rU_{\mathrm{old}}\right],\\
V_{\mathrm{anc}}&=U_{\mathrm{anc}}(x_{r,i},r,t)
 -(t-r)\operatorname{sg}\!\left[\widehat D_rU_{\mathrm{old}}\right],
\label{eq:practical-induced-velocities}
\end{align}
where $\operatorname{sg}$ denotes stop gradient.  Sharing makes $V_\theta-V_{\mathrm{old}}=U_\theta-U_{\mathrm{old}}$ and $V_\theta-V_{\mathrm{anc}}=U_\theta-U_{\mathrm{anc}}$, preventing a noisy finite-difference discrepancy from dominating either term. Following DiffusionNFT, define the implicit positive and negative predictors
\begin{equation}
V^+_\theta=(1-\beta)V_{\mathrm{old}}+\beta V_\theta,
\qquad
V^-_\theta=(1+\beta)V_{\mathrm{old}}-\beta V_\theta.
\label{eq:implicit-predictors}
\end{equation}
Let $\widehat x_{1,i}^{\pm}=x_{r,i}+(1-r)V_{\theta,i}^{\pm}$ and $a_i^\pm=\operatorname{sg}[\max\{\operatorname{mean}|\widehat x_{1,i}^{\pm}-x_{1,i}|,10^{-5}\}]$. The scalar loss minimized by CrystAF is
\begin{align}
\mathcal L_{\mathrm{CrystAF}}(\theta)
=\mathbb E_{g,i,\epsilon,j}\Bigg[&\frac{p_{gi}}{\beta}
\frac{\operatorname{MSE}_{\mathcal M}(\widehat x_{1,i}^{+},x_{1,i})}{a_i^+}
 +\frac{1-p_{gi}}{\beta}
\frac{\operatorname{MSE}_{\mathcal M}(\widehat x_{1,i}^{-},x_{1,i})}{a_i^-}
\nonumber\\
&+\lambda_{\mathrm{KL}}
 \operatorname{MSE}(V_\theta,V_{\mathrm{anc}})\Bigg].
\label{eq:complete-nft-loss}
\end{align}
The last term is the velocity-space surrogate used to anchor the policy to its frozen reference (reported as the KL regulariser); $\beta=0.1$ and $\lambda_{\mathrm{KL}}=10^{-4}$ for CrystAF. The reported configuration has unit advantage-clip scaling, so no additional multiplier appears above. Equations \ref{eq:reward-channels}--\ref{eq:complete-nft-loss} also make clear that the black-box scores affect optimisation only through $p_{gi}$: no gradient is taken through UMA, the clash test, or the volume calculation.

\subsection{Flow-map identity under the lower-endpoint convention}
\label{sec:flow-derivation}

We now derive Equation~\ref{eq:identity}.  Fix an ODE trajectory $\dot x_s=v(x_s,s)$ and its upper time $t$.  From Equation~\ref{eq:flow-map-definition},
\begin{equation}
(t-r)U(x_r,r,t)=\int_r^t v(x_s,s)\,ds=x_t-x_r.
\label{eq:displacement-identity}
\end{equation}
Move the lower endpoint along this same trajectory, so $d x_r/dr=v(x_r,r)$.  The chain rule gives
\begin{equation}
D_rU(x_r,r,t)
=\left.\frac{\partial U(x,r,t)}{\partial r}\right|_{x=x_r,t}
+\nabla_xU(x_r,r,t)\,v(x_r,r).
\label{eq:lower-total-derivative}
\end{equation}
Taking a total derivative of both sides of Equation~\ref{eq:displacement-identity} with respect to $r$ yields
\begin{equation}
-U(x_r,r,t)+(t-r)D_rU(x_r,r,t)=-v(x_r,r),
\end{equation}
and rearranging proves $v(x_r,r)=U(x_r,r,t)-(t-r)D_rU(x_r,r,t)$.  The spatial term in Equation~\ref{eq:lower-total-derivative} is therefore part of the identity; a partial derivative at fixed $x_r$ is not equivalent.

The exact expression is implicit because the direction of the spatial derivative contains the instantaneous velocity being recovered. As in MeanFlowNFT, the practical estimator uses the available conditional forward velocity $w_i$ from Equation~\ref{eq:forward-path}. With $t$ held fixed, both the lower time and its state are perturbed. To remain in the valid interval, set $r_+=\min(r+\delta,t)$, $r_-=\max(r-\delta,0)$, $\delta_+=r_+-r$, and $\delta_-=r-r_-$:
\begin{equation}
\widehat D_rU_{\mathrm{old}}
=\frac{
U_{\mathrm{old}}(x_{r,i}+\delta_+w_i,r_+,t)
-U_{\mathrm{old}}(x_{r,i}-\delta_-w_i,r_-,t)}
{r_+-r_-},\qquad \delta=0.005.
\label{eq:central-total-derivative}
\end{equation}
Away from the interval boundaries this is the symmetric central difference and has $O(\delta^2)$ truncation error. At a boundary it becomes a one-sided difference with $O(\delta)$ truncation error. This equation specifies the endpoint convention omitted from the compact main text: the upper endpoint $t$ is unchanged, while $(x_r,r)$ moves together along the forward conditional path. In contrast, the time-partial ablation uses the same time perturbations with $x_{r,i}$ unchanged in both evaluations.

\subsection{Definition of the flow-map-specific gradient diagnostic}
\label{sec:gradient-diagnostic}

Figure~\ref{fig:dose} compares two gradients on identical candidates, forward noise, model weights, and time pairs.  Let $\mathcal L_{\mathrm{MF}}$ be Equation~\ref{eq:complete-nft-loss}, and let $\mathcal L_{\mathrm{Diff}}$ be the same loss after replacing each induced predictor $V$ by its uncorrected map prediction $U$ (equivalently, deleting the $(t-r)\widehat D_rU$ term).  For a minibatch $B$, we first average each loss over the batch and then differentiate.  The plotted statistic is
\begin{equation}
G_B=100\,
\frac{\left\lVert
\nabla_\theta\mathcal L_{\mathrm{MF}}^{(B)}-
\nabla_\theta\mathcal L_{\mathrm{Diff}}^{(B)}
\right\rVert_2}
{\left\lVert\nabla_\theta\mathcal L_{\mathrm{MF}}^{(B)}\right\rVert_2+\epsilon_G}.
\label{eq:gradient-difference}
\end{equation}
Thus the reported percentage is a normed difference of batch gradients, not a projection and not the mean of per-sample ratios.  The $0.75\%$ reference-mix number is the batch-composition-weighted value for a 50--50 mixture of zero-length and adjacent $1/16$ intervals.  The bars are matched-batch diagnostics rather than estimates over independently trained seeds; no seed-level error bar is implied.  Repeating this diagnostic over independent batches is the appropriate way to attach sampling uncertainty without conflating it with training-seed variance.

\subsection{Induced-velocity estimator diagnostics}

We use the shared state--time finite difference in Equation~\ref{eq:central-total-derivative} to estimate $D_rU$ from the old policy and reuse it for the policy, old policy, and frozen anchor. Forward-mode AD would be preferable but is unavailable because the backbone's distance computation has no forward-mode rule. The correction of interest is only $3$--$6\%$ of $\lVert U\rVert$, while the finite-difference estimate is about six times larger and is dominated by the spatial term $(\nabla_x U)v$. Per sample, that term is nearly uncorrelated with the target ($\cos=-0.06$), whereas the time-partial component alone gives $+0.79$. Restricting the estimator to the time partial, however, damages the model in practice ($89.2\to74.1$), so the spatial term cannot simply be discarded.

\subsection{Training recipes and hyperparameters}
\label{sec:hp}

All optimization uses AdamW. Multi-GPU runs do not all-reduce: each rank trains on its own candidates and weights are averaged through the filesystem at the end of every epoch. All four ranks draw the same 20 families per epoch, so one epoch is 20 families with $4\times100$ candidates each. \texttt{w\_fmax} and \texttt{w\_stress} act as switches, not weights: a non-zero value turns a group-standardized term on and its magnitude is discarded, so the energy/force balance is set by $\lambda$ (\texttt{ef\_lambda}).

\paragraph{Starting points.} The distillation teacher is Clari-M after NFT post-training with a PoseBusters-ranked reward (Appendix~\ref{sec:si_distillation}); every CrystAF row inherits it, which is why all comparisons in the main text are made against CrystAF-base rather than against Clari-M. The released CrystAF-UMA weights were trained from CrystAF-base plus 12 epochs of PoseBusters-ranked NFT, whose checkpoint has since been deleted; that starting point scores $89.24\%$ PB at $\NFE{=}32$, level with CrystAF-base's $89.32\%$. All ablation arms, including a re-run of the full signal, start from CrystAF-base itself, so no PoseBusters-derived step separates them from the base row. The Clari-M row of Table~\ref{tab:generality} starts from the public checkpoint and has no PoseBusters step anywhere.

\begin{table}[H]
\caption{Post-training configurations. The CrystAF-UMA column is the released model; the ablation column applies to every arm of Table~\ref{tab:ablation}, which differ only as listed in Appendix~\ref{sec:ablation-details}.}
\label{tab:hp-stage2}
\begin{center}
{\scriptsize
\setlength{\tabcolsep}{3pt}
\begin{tabular}{lccc}
\toprule
 & CrystAF-UMA (released) & ablation arms & PB-rank reward \\
\midrule
Init & base + 12 ep.\ PB-rank NFT & CrystAF-base & base + 12 ep.\ PB-rank NFT \\
LoRA rank / $\alpha$ & $32$ / $64$ & $32$ / $64$ & $32$ / $64$ \\
Trainable fraction & $3.77\%$ & $3.77\%$ & $3.77\%$ \\
Rollout NFE / $\rho$ & $16$ / $0.75$ & $16$ / $0.75$ & $16$ / $0.75$ \\
Families per epoch & $20$ & $20$ & $20$ \\
Ranks $\times$ candidates per family & $4\times100$ & $4\times100$ & $4\times100$ \\
Epochs (read-out) & $6$ & $6$ & $12$ \\
Learning rate & $5\times10^{-6}$ & $5\times10^{-6}$ & $2\times10^{-5}$ \\
Weight decay / grad clip & $10^{-4}$ / $1.0$ & $10^{-4}$ / $1.0$ & $10^{-4}$ / $1.0$ \\
$(r,t)$ sampling & adjacent knots, $N{=}8$ & adjacent knots, $N{=}8$ & adjacent knots, $N{=}8$ \\
Central-difference $\varepsilon$ & $0.005$ & $0.005$ & $0.005$ \\
EMA decay & $0.999$ & $0.999$ & $0.999$ \\
$\beta$ / KL coefficient & $0.1$ / $10^{-4}$ & $0.1$ / $10^{-4}$ & $0.1$ / $10^{-4}$ \\
Advantage clip & $[-1,1]$ & $[-1,1]$ & $[-1,1]$ \\
Reward & $E$, $\bar F$, $F_{\max}$, $\sigma$, elig., cell & per arm & PB rank, clash veto, cell \\
$\lambda$ / $w_V$ & $0.5$ / $1.0$ & per arm & --- / $1.0$ \\
Seeds & $929$, $2029$ & $929$ & $929$, $2029$ \\
\bottomrule
\end{tabular}}
\end{center}
\end{table}

\subsection{Reward-path failure diagnostics}
\label{sec:rewardbugs}

Our first physical-reward implementation fell from PoseBusters $89.3\%$ to $41.5\%$ in three epochs, while the UMA-valid fraction dropped from $0.79$ to $0.002$ and mean cell error reached $416\%$. The failure came from two advantage construction bugs rather than the step size. First, the non-positive gate for invalid candidates was disabled whenever a target-driven term such as cell volume was active, allowing a physically rejected structure to receive maximum positive advantage. Second, maximum-force and stress retained a $-10^6$ sentinel for invalid candidates before group standardisation; a single sentinel then collapsed the valid entries onto nearly the same standardised value, effectively removing those channels. Enforcing the non-positive rejection gate and consistent invalid-value substitution fixes both effects and reaches $91.04\%$ at epoch 2 and $93.46\%$ at epoch 4 with the same recipe and epoch budget.

\subsection{Rollout-grid mismatch}
\label{sec:grid}

After the reward path was fixed, runs could still degrade after several epochs. Training rollouts used a 16-NFE grid ($\rho=0.75$) inherited from distillation, while evaluation used 32 NFE ($\rho=1$). The coarser rollout grid produced more geometry-filter failures and therefore a sparser, noisier learning signal. Rolling out on the reported grid keeps the UMA-valid fraction near $0.90$ instead of falling toward $0.70$, lowers epoch-7 KL to the frozen reference by about $45\times$, and extends the usable training window from roughly 6 epochs to 13. At epoch 12 it reaches volume error $1.81\%$ and PDD $10.14$, while validity plateaus near $93\%$.

\subsection{Physical-reward training dynamics}

\begin{figure}[H]
\centering
\begin{subfigure}[b]{0.32\textwidth}
\includegraphics[width=\textwidth]{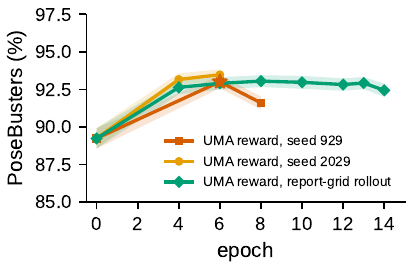}
\caption{}
\end{subfigure}
\begin{subfigure}[b]{0.32\textwidth}
\includegraphics[width=\textwidth]{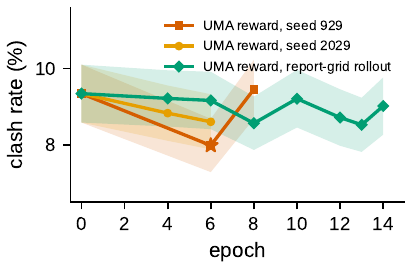}
\caption{}
\end{subfigure}
\begin{subfigure}[b]{0.32\textwidth}
\includegraphics[width=\textwidth]{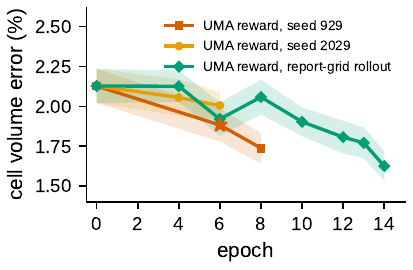}
\caption{}
\end{subfigure}

\begin{subfigure}[b]{0.32\textwidth}
\includegraphics[width=\textwidth]{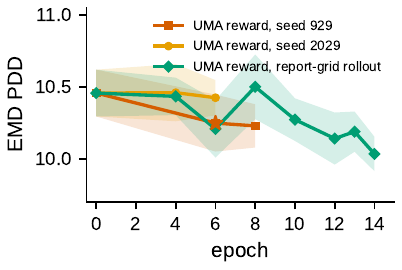}
\caption{}
\end{subfigure}
\begin{subfigure}[b]{0.32\textwidth}
\includegraphics[width=\textwidth]{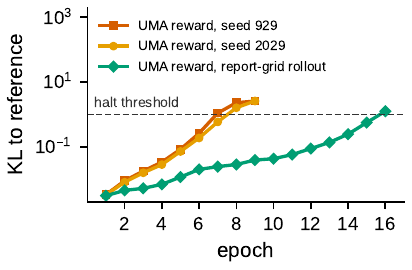}
\caption{}
\end{subfigure}
\begin{subfigure}[b]{0.32\textwidth}
\includegraphics[width=\textwidth]{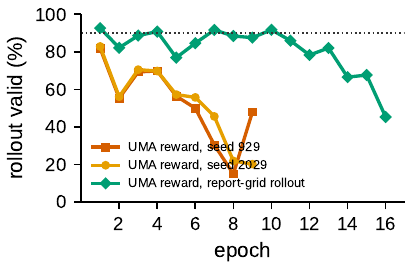}
\caption{}
\end{subfigure}
\caption{Physical-reward training dynamics at 32 NFE. (a)--(d) Validity, clash, volume error, and PDD for two seeds and a report-grid rollout; stars mark reported checkpoints. (e) KL to the frozen reference. (f) Fraction of rollout candidates passing the geometry filter.}
\label{fig:training}
\end{figure}

\subsection{The PoseBusters-reward training curves}

\begin{figure}[H]
\centering
\begin{subfigure}[b]{0.32\textwidth}
\includegraphics[width=\textwidth]{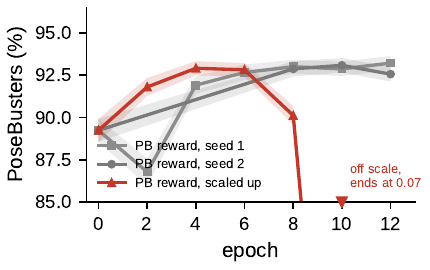}
\caption{}
\end{subfigure}
\begin{subfigure}[b]{0.32\textwidth}
\includegraphics[width=\textwidth]{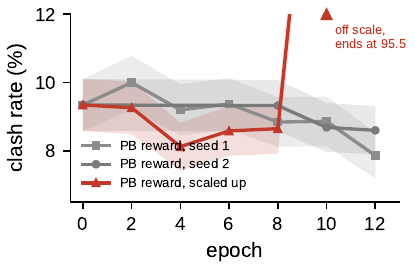}
\caption{}
\end{subfigure}
\begin{subfigure}[b]{0.32\textwidth}
\includegraphics[width=\textwidth]{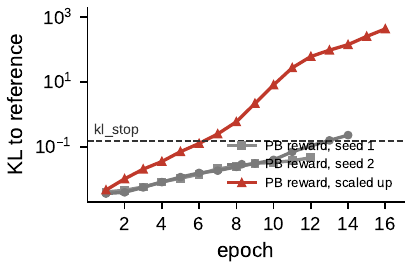}
\caption{}
\end{subfigure}
\caption{PoseBusters-reward training dynamics. (a) Report-grid pass rate and (b) clash rate against epoch for two seeds and the scaled-up run. (c) KL to the frozen reference on a log axis.}
\label{fig:si_pbtrain}
\end{figure}

\subsection{Scale-up: more families and a larger group}

The scaled-up run differs from the report-grid rollout of Appendix~\ref{sec:grid} in exactly two fields: families per epoch $80 \to 240$ and candidates per family $100 \to 150$ (8 ranks instead of 4). That rollout run is therefore the matched control, and because the two differ in families per epoch, epochs are not comparable between them --- families \emph{seen} is.

\begin{table}[H]
\caption{Scaling results for the physical-reward rollout at $\NFE{=}32$. Rows are aligned on families seen rather than epochs; the scaled-up run and its 80-family control are shown at matched family counts, with an additional scaled-up point at 1920 families. \textbf{Bold} is best per column.}
\label{tab:scaleup}
\begin{center}
{\small
\setlength{\tabcolsep}{4pt}
\begin{tabular}{lrccccc}
\toprule
& families & epoch & PB \% $\uparrow$ & clash \% $\downarrow$ & Vol.Err $\downarrow$ & EMD PDD $\downarrow$ \\
\midrule
\multicolumn{7}{l}{\textit{240 families/epoch, group 150, 8 ranks}} \\
scaled up & 480 & 2 & 92.92 \small{$\pm$0.46} & 9.24 \small{$\pm$0.75} & 2.07 \small{$\pm$0.11} & 10.39 \small{$\pm$0.18} \\
scaled up & 960 & 4 & \best{93.29} \small{$\pm$0.36} & 8.73 \small{$\pm$0.73} & 1.83 \small{$\pm$0.10} & 10.17 \small{$\pm$0.15} \\
scaled up & 1440 & 6 & 93.17 \small{$\pm$0.40} & 8.33 \small{$\pm$0.69} & 1.87 \small{$\pm$0.10} & 10.26 \small{$\pm$0.12} \\
scaled up & 1920 & 8 & 83.22 \small{$\pm$0.71} & 22.83 \small{$\pm$1.10} & 3.60 \small{$\pm$0.14} & 12.21 \small{$\pm$0.17} \\
\midrule
\multicolumn{7}{l}{\textit{80 families/epoch, group 100, 4 ranks --- the matched control}} \\
control & 480 & 6 & 92.91 \small{$\pm$0.41} & 9.16 \small{$\pm$0.75} & 1.92 \small{$\pm$0.10} & 10.21 \small{$\pm$0.20} \\
control & 640 & 8 & 93.05 \small{$\pm$0.40} & 8.57 \small{$\pm$0.70} & 2.06 \small{$\pm$0.11} & 10.50 \small{$\pm$0.22} \\
control & 960 & 12 & 92.82 \small{$\pm$0.42} & 8.72 \small{$\pm$0.74} & 1.81 \small{$\pm$0.10} & 10.14 \small{$\pm$0.18} \\
control & 1120 & 14 & 92.44 \small{$\pm$0.44} & 9.02 \small{$\pm$0.75} & \best{1.63} \small{$\pm$0.09} & \best{10.03} \small{$\pm$0.12} \\
\bottomrule
\end{tabular}}
\end{center}
\end{table}

\begin{figure}[H]
\centering
\begin{subfigure}[b]{0.42\textwidth}
\includegraphics[width=\textwidth]{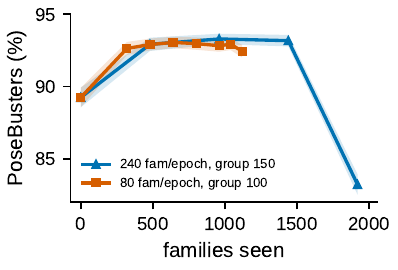}
\caption{}
\end{subfigure}
\hfill
\begin{subfigure}[b]{0.42\textwidth}
\includegraphics[width=\textwidth]{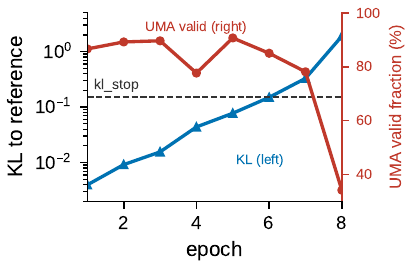}
\caption{}
\end{subfigure}
\caption{Scaled-up physical-reward rollout and matched control. (a) Report-grid PoseBusters against \emph{families seen}. (b) KL to the frozen reference and UMA-valid rollout fraction against epoch for the scaled-up run.}
\label{fig:si_scaleup}
\end{figure}

Training-side signals across the eight epochs: epoch-mean KL $0.0040$, $0.0092$, $0.0156$, $0.0438$, $0.0774$, $0.1496$, $0.3291$, $1.8963$, against a UMA valid fraction of $0.867$, $0.892$, $0.897$, $0.784$, $0.907$, $0.851$, $0.788$, $0.357$. The two only move together at the very end: the first seven epochs drift steadily while validity stays near $0.85$, and then the eighth loses two thirds of its candidates to the geometry filter in one epoch.

\subsection{Second seeds}

\begin{table}[H]
\caption{Seed-to-seed variation at $\NFE{=}32$, 200 families $\times$ 20 samples. Both the released CrystAF-UMA model and the PoseBusters-ranked reward start from the same warm-started checkpoint (Appendix~\ref{sec:hp}); the report-grid rows roll out on the evaluation grid. PB-rank reward has $7.86$ clash and $1.61$ volume error for one seed, and $8.60$ and $1.88$ for the second.}
\label{tab:seeds}
\begin{center}
{\small
\setlength{\tabcolsep}{4pt}
\begin{tabular}{lccccc}
\toprule
& epoch & PB \% $\uparrow$ & clash \% $\downarrow$ & Vol.Err $\downarrow$ & EMD PDD $\downarrow$ \\
\midrule
CrystAF-UMA, seed 929 & 6 & 93.01 \small{$\pm$0.43} & 7.98 \small{$\pm$0.70} & 1.88 \small{$\pm$0.10} & 10.25 \small{$\pm$0.20} \\
CrystAF-UMA, seed 2029 & 6 & 93.48 \small{$\pm$0.36} & 8.60 \small{$\pm$0.72} & 2.01 \small{$\pm$0.08} & 10.42 \small{$\pm$0.12} \\
CrystAF-UMA, report grid & 12 & 92.82 \small{$\pm$0.42} & 8.72 \small{$\pm$0.71} & 1.81 \small{$\pm$0.10} & 10.14 \small{$\pm$0.17} \\
CrystAF-UMA, report grid & 14 & 92.44 \small{$\pm$0.44} & 9.02 \small{$\pm$0.74} & 1.63 \small{$\pm$0.09} & 10.03 \small{$\pm$0.16} \\
\midrule
PB-rank reward, seed 929 & 12 & 93.21 \small{$\pm$0.42} & 7.86 \small{$\pm$0.67} & 1.61 \small{$\pm$0.11} & 10.07 \small{$\pm$0.23} \\
PB-rank reward, seed 2029 & 12 & 92.56 \small{$\pm$0.43} & 8.60 \small{$\pm$0.71} & 1.88 \small{$\pm$0.09} & 10.12 \small{$\pm$0.14} \\
\midrule
\multicolumn{6}{l}{\textit{defensible range across seeds}} \\
CrystAF-UMA & --- & 93.0--93.5 & 8.0--8.6 & 1.9--2.0 & 10.3--10.4 \\
PB-rank reward & --- & 92.6--93.2 & 7.9--8.6 & 1.6--1.9 & 10.1 \\
\bottomrule
\end{tabular}}
\end{center}
\end{table}

\subsection{Cell volume before and after reinforcement learning}

\begin{figure}[H]
\centering
\includegraphics[width=0.52\textwidth]{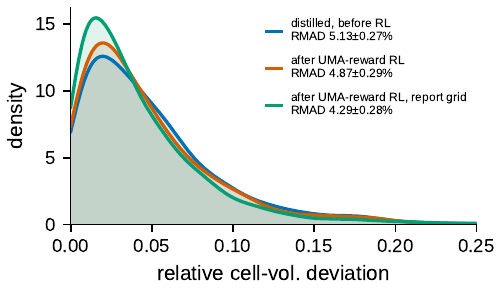}
\caption{Relative cell-volume deviation distributions for generated structures. The plot contains 4000 structures before and after physical-reward RL and after report-grid rollout. RMAD is the mean absolute relative deviation, reported as mean $\pm$ s.d.\ across 20 samples.}
\label{fig:si_volkde}
\end{figure}

\subsection*{D.\quad Training-free and training-time baselines}

\subsection{Training-free baselines: implementation and settings}
\label{sec:trainfree-details}

All training-free methods run inside the sampler, on the same CrystAF weights and the same random seeds as the uncorrected rows. UMA-OMC (\texttt{uma-s-1p1}) is evaluated in float32 with autograd forces, outside the sampler's inference mode and bfloat16 autocast.

\paragraph{Force guidance.} At a jump $r\to t$ with $r\ge r_0$ we form $\hat x_1=x_r+(1-r)U_\theta(x_r,r,t)$, compute UMA forces $F$ on $\hat x_1$, and set $\Delta=\eta F$, rescaled per structure so that no atom moves more than $\delta$. The step is re-aimed, $x_t=x_r+\frac{t-r}{1-r}(\hat x_1+\Delta-x_r)$, so an early correction is applied partially and the last one in full. We report two settings chosen before seeing results: a gentle one ($\eta=0.01$\,\AA$^2$/eV, $\delta=0.05$\,\AA) and a strong one ($\eta=0.05$, $\delta=0.2$\,\AA), both with $r_0=0.5$, i.e.\ 16 guided jumps out of 32. Only atomic positions are guided; the cell is left to the model.

\paragraph{Relaxation.} After sampling, each structure takes $K\in\{10,50\}$ steps $x\leftarrow x+0.05\,F/\lVert F\rVert_\infty$ (\AA), with the cell fixed. We fix the cell because UMA's equilibrium cell is larger than the experimental one: on 36 generated structures, relaxing the cell moved the signed volume error from $+4.16\%$ to $+6.50\%$, while fixed-cell relaxation removed all clashes and left the volume unchanged.

\paragraph{Classical chain.} Parity inversion flips whole molecules whose stereocentres disagree with the R/S labels of the input graph and then reflects remaining branches; restrained MMFF94 (force constant 50, at most 0.15\,\AA{} displacement, 400 iterations) repairs local strain; a rigid-body declash (60 iterations) removes intermolecular contacts without changing molecular geometry; cell calibration multiplies the volume by $0.985$, the median ratio measured on the training split. None of these calls UMA.

\paragraph{Constraint projection.} The PCFM-style baseline projects bond lengths onto RDKit distance-geometry bounds after sampling. It runs 6 closed-form Gauss--Newton iterations: each violated bond moves both atoms along the bond by half the correction, overlapping bonds are averaged per atom, and no atom moves more than $0.25$\,\AA{}. The chirality residual of the original PCFM is left out. With it, projection reaches $99.97\%$ stereochemical agreement but costs 15 PoseBusters points (Table~\ref{tab:recipes}).

\paragraph{How many knobs.} Force guidance has three (scale, cap, start time), relaxation three (steps, step size, cell), constraint projection two (iterations, cap), and the classical chain five (MMFF force constant, displacement cap, iterations, declash iterations, calibration factor). Post-training has its own hyperparameters (Table~\ref{tab:hp-stage2}), but they are set once during training and cost nothing when sampling.

\subsection{The rest of the corrector chain}

\begin{table}[H]
\caption{Results for sampling-time correction-chain variants. These rows extend Table~\ref{tab:main}; the chain \emph{overwrites} rather than compounds. The CrystAF-base block applies the identical chain and cell calibration to the weights before reinforcement learning. \textbf{Bold} is best per column.}
\label{tab:stereo}
\begin{center}
{\small
\setlength{\tabcolsep}{4pt}
\begin{tabular}{lcccccc}
\toprule
& NFE & PB \% $\uparrow$ & clash \% $\downarrow$ & Vol.Err $\downarrow$ & EMD PDD $\downarrow$ & stereo \% $\uparrow$ \\
\midrule
\multicolumn{7}{l}{\textit{CrystAF-UMA weights, no corrections}} \\
no corrections & 32 & 93.01 \small{$\pm$0.43} & 7.98 \small{$\pm$0.70} & 1.88 \small{$\pm$0.10} & 10.25 \small{$\pm$0.20} & 48.38 \\
no corrections & 50 & \best{94.05} \small{$\pm$0.34} & 8.13 \small{$\pm$0.71} & 1.84 \small{$\pm$0.12} & \best{10.24} \small{$\pm$0.29} & 50.47 \\
\midrule
\multicolumn{7}{l}{\textit{+ parity inversion, constrained MMFF, rigid declash}} \\
+ corrections & 8 & 92.31 \small{$\pm$0.41} & 7.32 \small{$\pm$0.73} & 3.79 \small{$\pm$0.19} & 13.85 \small{$\pm$0.20} & 94.84 \\
+ corrections & 16 & 91.66 \small{$\pm$0.42} & 3.13 \small{$\pm$0.49} & 2.21 \small{$\pm$0.13} & 11.27 \small{$\pm$0.29} & \best{95.53} \\
+ corrections & 32 & 93.43 \small{$\pm$0.33} & 2.80 \small{$\pm$0.47} & 1.93 \small{$\pm$0.10} & 10.78 \small{$\pm$0.16} & 95.09 \\
+ corrections & 50 & 93.78 \small{$\pm$0.36} & 2.20 \small{$\pm$0.43} & 1.80 \small{$\pm$0.11} & 10.69 \small{$\pm$0.25} & 95.34 \\
\ \ + cell calibration & 32 & 93.42 \small{$\pm$0.36} & 2.55 \small{$\pm$0.46} & 1.58 \small{$\pm$0.09} & 10.50 \small{$\pm$0.15} & 95.43 \\
\ \ + cell calibration & 50 & 93.57 \small{$\pm$0.36} & 2.80 \small{$\pm$0.46} & 1.64 \small{$\pm$0.12} & 10.61 \small{$\pm$0.29} & 95.34 \\
\midrule
\multicolumn{7}{l}{\textit{the same chain on the PB-rank reward weights}} \\
+ corrections & 32 & \best{93.73} \small{$\pm$0.37} & 2.58 \small{$\pm$0.44} & 1.64 \small{$\pm$0.09} & 10.42 \small{$\pm$0.16} & 95.26 \\
\ \ + cell calibration & 32 & 93.58 \small{$\pm$0.37} & \best{2.12} \small{$\pm$0.42} & \best{1.51} \small{$\pm$0.09} & 10.47 \small{$\pm$0.19} & 95.20 \\
\midrule
\multicolumn{7}{l}{\textit{the same chain on the CrystAF-base weights (no RL)}} \\
+ corrections & 32 & 92.91 \small{$\pm$0.41} & 2.64 \small{$\pm$0.46} & 1.89 \small{$\pm$0.10} & 10.64 \small{$\pm$0.17} & 94.92 \\
+ corrections & 50 & 92.81 \small{$\pm$0.44} & 3.08 \small{$\pm$0.50} & 1.87 \small{$\pm$0.14} & 10.88 \small{$\pm$0.46} & 95.28 \\
\ \ + cell calibration & 32 & 92.47 \small{$\pm$0.43} & 2.66 \small{$\pm$0.45} & 1.56 \small{$\pm$0.11} & 10.53 \small{$\pm$0.25} & 95.17 \\
\ \ + cell calibration & 50 & 92.99 \small{$\pm$0.45} & 2.88 \small{$\pm$0.46} & 1.53 \small{$\pm$0.13} & 10.57 \small{$\pm$0.36} & 95.17 \\
\bottomrule
\end{tabular}}
\end{center}
\end{table}

The $\NFE{=}32$ rows of the chain on the CrystAF-UMA and CrystAF-base weights are also in Table~\ref{tab:main}. EMD PDD is the one column that is independent in every row here, and it gets \emph{worse} under the chain ($10.24 \to 10.69$ at $\NFE{=}50$) --- the price of the local moves.

The CrystAF-base block separates what reinforcement learning contributes once the corrections are applied. Against the matching CrystAF-UMA rows above, the RL weights stay ahead on PoseBusters by $0.5$--$1.0$ points ($93.43$ vs.\ $92.91$ and $93.78$ vs.\ $92.81$ without calibration, $93.42$ vs.\ $92.47$ and $93.57$ vs.\ $92.99$ with it), i.e.\ $1.0$--$1.7$ combined standard errors and comparable to the $\pm0.5$ run-to-run spread; clash, volume error, PDD and stereochemistry are level. Without corrections the same comparison is $+3.7$ points at $\NFE{=}32$ ($93.01$ vs.\ $89.32$). Most of the validity gain from RL is therefore also recoverable by the sampling-time chain, and what survives correction is a small but consistent PoseBusters margin. On the base weights the chain also costs PDD, as it does on the RL weights ($10.44 \to 10.64$ at $\NFE{=}32$).

\subsection{Sampling-time recipes}

\begin{table}[h]
\caption{Results for sampling-time recipes on the distilled student CrystAF-base. Rows use the same weights, $\NFE{=}16$ unless noted, and involve no training.}
\label{tab:recipes}
\centering
\small
\begin{tabular}{lrrrrr}
\toprule
recipe & stereo \% & PB \% & clash \% & Vol.Err & EMD PDD \\
\midrule
none & 49.91 & 85.49 & 13.85 & 2.09 & 10.92 \\
parity inversion & 85.19 & 85.44 & 19.10 & $\approx$2.09 & --- \\
\ \ + MMFF & 95.08 & 91.71 & 23.20 & 2.02 & 11.05 \\
\ \ + MMFF + declash & 95.24 & 92.16 & 2.27 & 2.06 & 11.03 \\
\ \ same, $\NFE{=}50$ & 95.59 & 92.63 & 2.40 & 1.78 & 10.57 \\
constraint projection & 99.97 & 70.13 & 14.14 & 2.06 & 10.79 \\
declash + cell scale ($\rho{=}0.50$) & 84.97 & 93.41 & 1.12 & 1.65 & 10.67 \\
\bottomrule
\end{tabular}
\end{table}

\subsection{Relax-and-distill: settings}
\label{sec:rd-details}

The training-time baseline (\texttt{crystaf\_physdistill\_base}) starts from CrystAF-base and uses the same LoRA. For each of 20 families per epoch and rank, it samples 24 candidates with the EMA policy at $\NFE{=}32$, $\rho=1$. Each candidate is relaxed for 25 UMA steps (largest force component moved $0.05$\,\AA{} per step, fixed cell, no lattice rescale), and the flow map is regressed onto the relaxed structures for two inner epochs (batch 8, learning rate $10^{-5}$, EMA $0.999$). It runs for six epochs on four ranks. No reward, ranking, reference policy, or PoseBusters term is involved. The relaxation removes most close contacts: on 36 generated structures, 25 steps took the clash fraction from $0.056$ to $0$ and the mean maximum force from $12.5$ to $3.4$\,eV\,\AA$^{-1}$, with the cell unchanged.

\subsection*{E.\quad Signal ablations}

\subsection{Signal and algorithm ablations}
\label{sec:ablation-details}

Every arm in Table~\ref{tab:ablation} starts from CrystAF-base and runs the CrystAF-UMA recipe of Table~\ref{tab:hp-stage2} for six epochs: same families, candidates per family, learning rate, EMA, rollout grid, seed, and read-out epoch. The arms differ in exactly the fields below; the config files are in the repository.

\begin{center}
\small
\begin{tabular}{ll}
\toprule
arm & change against the full signal \\
\midrule
full signal & none ($\lambda=0.5$; $F_{\max}$ and $\sigma$ on; $w_V=1$) \\
$E,F,\sigma$ & $w_V=0$ \\
$E$ only & $\lambda=1$; $F_{\max}$, $\sigma$ off; $w_V=0$ \\
feasibility only & energy and force channels off; $F_{\max}$, $\sigma$ off; $w_V=1$ \\
eligibility gate only & all UMA channels off; $w_V=0$ (4 epochs) \\
DiffusionNFT on $U$ & update applied to $U$ directly (no Equation~\ref{eq:identity}) \\
\bottomrule
\end{tabular}
\end{center}

With the volume term off, an ineligible candidate receives the most negative advantage $-1$; with it on, it is floored at $0$ (Equation~\ref{eq:rejection-consistent-advantage}). The eligibility gate is part of every arm, including feasibility only, because without it there is no way to reject a structure the potential cannot score. The PoseBusters-ranked arm in Table~\ref{tab:ablation} is an earlier run with a different starting point (Appendix~\ref{sec:hp}); it is reported for reference and not as a matched control.

\subsection{Training-side signals and the read-out epoch}
\label{sec:train-signals}

Table~\ref{tab:train-signals} lists, for every matched arm, the fraction of rollout candidates that pass the eligibility check and the epoch-mean KL surrogate to the frozen reference. Both are visible during training, without any evaluation. The arms that use the flow-map conversion enter the same drift regime as the released run of Appendix~\ref{sec:grid}, but about one epoch earlier: KL crosses $0.15$ in epoch 4 or 5, and the valid fraction falls in epoch 5 or 6. We declared before any evaluation that every arm would be read out at both epoch 4 and epoch 6. The main text uses epoch 4 for all arms. Table~\ref{tab:ablation-ep6} gives epoch 6. DiffusionNFT on $U$ drifts much more slowly, which Section~\ref{sec:what-learned} discusses.

\begin{table}[H]
\caption{Rollout valid fraction and epoch-mean KL per epoch (rank 0), matched arms from CrystAF-base.}
\label{tab:train-signals}
\centering
\scriptsize
\setlength{\tabcolsep}{2.5pt}
\begin{tabular}{lcccccc|cccccc}
\toprule
& \multicolumn{6}{c|}{valid fraction, epoch 1--6} & \multicolumn{6}{c}{KL, epoch 1--6} \\
\midrule
full signal & 0.70 & 0.41 & 0.56 & 0.62 & 0.39 & 0.14 & 0.001 & 0.007 & 0.018 & 0.043 & 0.235 & 1.127 \\
$E,F,\sigma$ & 0.69 & 0.39 & 0.51 & 0.56 & 0.33 & 0.35 & 0.001 & 0.011 & 0.040 & 0.178 & 0.940 & 1.888 \\
$E$ only & 0.69 & 0.37 & 0.46 & 0.41 & 0.05 & 0.29 & 0.001 & 0.018 & 0.089 & 0.696 & 1.973 & 2.397 \\
feasibility only & 0.68 & 0.37 & 0.54 & 0.57 & 0.29 & 0.20 & 0.001 & 0.010 & 0.035 & 0.145 & 0.985 & 2.034 \\
eligibility gate only & 0.70 & 0.39 & 0.54 & 0.62 & --- & --- & 0.001 & 0.007 & 0.023 & 0.057 & --- & --- \\
DiffusionNFT on $U$ & 0.71 & 0.55 & 0.73 & 0.76 & 0.61 & 0.68 & 0.000 & 0.002 & 0.004 & 0.012 & 0.038 & 0.128 \\
\bottomrule
\end{tabular}
\end{table}

\subsection{The potential's view of the samples}
\label{sec:uma-diag}

Table~\ref{tab:uma-diag} scores every generated structure with UMA single points, the same quantities the reward uses. $\Delta E$ is the change in mean energy per molecule relative to CrystAF-base, averaged over families, so it compares the same molecules. Post-training does not lower the potential's energy. Every arm trained with our update, including the one rewarded on energy, ends \emph{higher} in energy and stress than CrystAF-base. Its cells also end closer to the experimental volume. UMA's equilibrium cell is larger than the experimental one (Appendix~\ref{sec:trainfree-details}), so a cell that moves toward experiment looks compressed to an unrelaxed single point. The advantages are also group-relative. They rank candidates within a family and cannot hold the absolute energy level fixed, so the level drifts even for the gate-only arm. DiffusionNFT on $U$ goes the other way, with lower stress and a larger volume error. Relaxation is the only route that lowers the energy substantially.

\begin{table}[H]
\caption{UMA single-point diagnostics of generated structures (200 families $\times$ 20, NFE 32; post-trained arms at epoch 4). Eligible: passes the $0.8$\,\AA{} distance and density check. Median $|\Delta V/V|$ is against the experimental cell.}
\label{tab:uma-diag}
\centering
\small
\resizebox{\textwidth}{!}{%
\begin{tabular}{lccccc}
\toprule
method & eligible & $\Delta E$ / molecule (eV) & median $F_{\max}$ (eV/\AA) & median $\lVert\sigma\rVert$ & median $|\Delta V/V|$ \\
\midrule
CrystAF-base & 0.853 & $+0.00$ & 8.59 & 0.058 & 0.0386 \\
post-training, full signal & 0.819 & $+0.91$ & 11.07 & 0.079 & 0.0331 \\
post-training, $E,F,\sigma$ & 0.779 & $+1.60$ & 11.65 & 0.095 & 0.0313 \\
post-training, gate only & 0.836 & $+0.77$ & 10.65 & 0.075 & 0.0359 \\
post-training, feasibility only & 0.805 & $+1.42$ & 11.60 & 0.089 & 0.0315 \\
DiffusionNFT on $U$ & 0.886 & $-0.27$ & 8.80 & 0.043 & 0.0480 \\
relax-and-distill & 0.807 & $+0.39$ & 9.28 & 0.069 & 0.0340 \\
base + UMA relaxation, 10 steps & 0.969 & $-3.57$ & 4.06 & 0.018 & 0.0375 \\
base + PCFM bond projection & 0.856 & $+0.01$ & 8.54 & 0.059 & 0.0391 \\
\bottomrule
\end{tabular}}
\end{table}

\begin{table}[H]
\caption{The matched arms of Table~\ref{tab:ablation} read out at epoch 6. All of them except DiffusionNFT on $U$ have passed the drift point flagged by the training-side signals (Table~\ref{tab:train-signals}).}
\label{tab:ablation-ep6}
\centering
\small
\setlength{\tabcolsep}{2.5pt}
\begin{tabular}{lcccc}
\toprule
Arm (epoch 6) & PB \% $\uparrow$ & clash \% $\downarrow$ & Vol.Err $\downarrow$ & EMD PDD $\downarrow$ \\
\midrule
full signal: $E$, $F$, $\sigma$, eligibility, cell volume & 89.41 \small{$\pm$0.50} & 10.60 \small{$\pm$0.78} & 1.57 \small{$\pm$0.09} & 9.94 \small{$\pm$0.13} \\
$E$, $F$, $\sigma$, eligibility & 87.91 \small{$\pm$0.55} & 10.64 \small{$\pm$0.81} & 1.65 \small{$\pm$0.09} & 10.01 \small{$\pm$0.11} \\
$E$, eligibility & 87.47 \small{$\pm$0.58} & 10.06 \small{$\pm$0.79} & 1.63 \small{$\pm$0.08} & 10.00 \small{$\pm$0.11} \\
feasibility only: eligibility, cell volume (no UMA) & 88.17 \small{$\pm$0.57} & 10.71 \small{$\pm$0.81} & 1.68 \small{$\pm$0.08} & 10.00 \small{$\pm$0.10} \\
DiffusionNFT on $U$, full signal & 93.07 \small{$\pm$0.38} & 10.41 \small{$\pm$0.82} & 2.90 \small{$\pm$0.12} & 11.09 \small{$\pm$0.16} \\
relax-and-distill (training-time) & 86.25 \small{$\pm$0.76} & 12.29 \small{$\pm$0.85} & 1.69 \small{$\pm$0.10} & 10.16 \small{$\pm$0.16} \\
\bottomrule
\end{tabular}
\end{table}

\subsection*{F.\quad Other generators and crystal recovery}

\subsection{Generality of the objective}

\begin{figure}[h]
\centering

\begin{subfigure}[b]{0.32\textwidth}
    \centering
    \includegraphics[width=\textwidth]{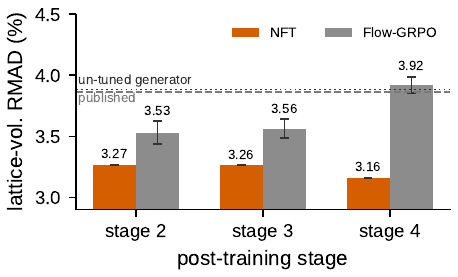}
    \caption{}
    \label{fig:si_grpo}
\end{subfigure}
\hfill
\begin{subfigure}[b]{0.32\textwidth}
    \centering
    \includegraphics[width=\textwidth]{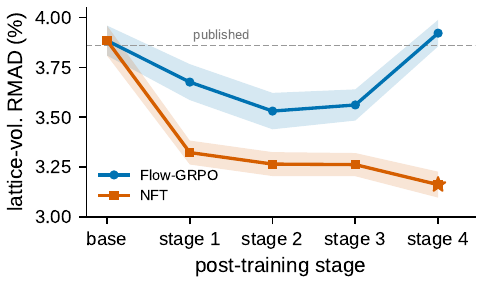}
    \caption{}
    \label{fig:si_traj}
\end{subfigure}
\hfill
\begin{subfigure}[b]{0.32\textwidth}
    \centering
    \includegraphics[width=\textwidth]{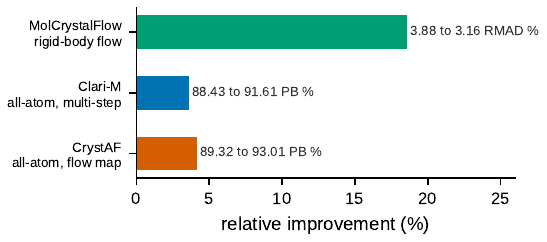}
    \caption{}
    \label{fig:si_generality}
\end{subfigure}

\caption{Generality diagnostics for NFT and Flow-GRPO. (a) Stage-wise values on MolCrystalFlow with wall-clock-matched stages and standard-deviation error bars. (b) Trajectories over the post-training schedule. (c) Relative changes across generators, with native headline metrics printed beside each bar.}
\label{fig:si_generality_all}
\end{figure}

\subsection{The same post-training route on other generators}
\label{sec:generality-details}
\label{sec:generality}

The comparison above uses a single few-step generator, so we also apply UMA feedback to two others, each on its own protocol (Table~\ref{tab:generality}). On the public Clari-M checkpoint, which has no PoseBusters-derived step anywhere in its history, two epochs raise PB from $88.4\%$ to $91.6\%$ on 1000 validation families. On MolCrystalFlow, which moves rigid molecules, the cell-volume error falls from $3.88\%$ to $3.16\%$, and with reward and budget matched, NFT beats Flow-GRPO \citep{liu2025flowgrpo} at every stage ($3.16\%$ against $3.92\%$).

\begin{table}[H]
\caption{The same post-training route on other generators, each on its native protocol. (a) UMA energy, force, and stress with a representation-specific feasibility term. Clari-M starts from the public checkpoint and is evaluated on 1000 validation families. (b) NFT and Flow-GRPO on MolCrystalFlow with matched reward, schedule, and stage budget.}
\label{tab:generality}
\centering
\small
\setlength{\tabcolsep}{3pt}
\begin{tabular}{llccc}
\toprule
\multicolumn{5}{l}{\textit{(a) Other generators}} \\
Generator (state) & Reward & Metric & base & $+$ post-training \\
\midrule
MolCrystalFlow (rigid bodies) & UMA $E$, $F$, $\sigma$, cell & Vol.\ RMAD \% $\downarrow$ & 3.88$\pm$0.08 & \best{3.16}$\pm$0.06 \\
Clari-M (all atom, 99 NFE) & UMA $E$, $F$, $\sigma$, eligibility & PB \% $\uparrow$ & 88.43$\pm$0.33 & \best{91.61}$\pm$0.28 \\
\midrule
\multicolumn{5}{l}{\textit{(b) NFT vs.\ Flow-GRPO (MolCrystalFlow, Vol.\ RMAD \% $\downarrow$)}} \\
Stage & wall-clock, NFT / GRPO & NFT & Flow-GRPO & $\Delta$ \\
\midrule
2 & 0.97\,h / 0.71\,h & \best{3.27}$\pm$0.06 & 3.53$\pm$0.09 & $+0.27$ \\
3 & 0.29\,h / 0.29\,h & \best{3.26}$\pm$0.06 & 3.56$\pm$0.08 & $+0.30$ \\
4 & 0.35\,h / 0.35\,h & \best{3.16}$\pm$0.06 & 3.92$\pm$0.07 & $+0.76$ \\
\bottomrule
\end{tabular}
\end{table}

\subsection{Structure recovery at other candidate budgets}
\label{sec:recovery-ns30}

Table~\ref{tab:solved-ns400} gives the benchmark with 400 candidates and $k=200$, and Table~\ref{tab:solved-ns30} with $n_s=k=30$, i.e.\ without energy ranking. Both use the same models and samplers as Table~\ref{tab:solved}. With only 30 candidates, most OXtal targets are either always or never solved. The CSP subsets move in steps of one target, and on CSP5 only two targets are solvable by any model. The teaching set is the only column with enough targets to compare methods. On it, every CrystAF row is above Clari-M ($0.411$--$0.427$ vs.\ $0.397$), and CrystAF-UMA and CrystAF-base are level. Without energy ranking the chain no longer helps: it leaves the teaching score at $0.411$ on CrystAF-base and $0.423$ on CrystAF-UMA, against $0.423$ and $0.427$ without it. Its gain at $k=30$ of 150 in Table~\ref{tab:solved} therefore depends on the UMA ranking picking out the repaired candidates. With 400 candidates the chain helps both models again. On CrystAF-base it raises every column (teaching $0.652\to0.667$, flexible $0.469\to0.499$, CSP7 $0.528\to0.595$), and CrystAF-base + chain is the best row on every subset except flexible, where CrystAF-UMA + chain leads ($0.522$).

\begin{table}[H]
\caption{Structure recovery with $n_s=400$, $k=200$ (bootstrap mean $\pm$ SE over samples). Same models, samplers, and COMPACK settings as Table~\ref{tab:solved}.}
\label{tab:solved-ns400}
\centering
\small
\setlength{\tabcolsep}{3.5pt}
\resizebox{\textwidth}{!}{%
\begin{tabular}{lcccccc}
\toprule
& Rigid (50) & Flexible (50) & CSP5 (6) & CSP6 (5) & CSP7 (8) & Teaching (779) \\
\midrule
Clari-M & $0.825{\pm}0.015$ & $0.422{\pm}0.023$ & $0.644{\pm}0.058$ & $0.564{\pm}0.078$ & $0.408{\pm}0.086$ & $0.641{\pm}0.006$ \\
CrystAF-base & $0.850{\pm}0.020$ & $0.469{\pm}0.021$ & $0.713{\pm}0.112$ & $0.625{\pm}0.152$ & $0.528{\pm}0.090$ & $0.652{\pm}0.006$ \\
CrystAF-base + chain & $0.859{\pm}0.017$ & $0.499{\pm}0.028$ & $0.872{\pm}0.117$ & $0.697{\pm}0.121$ & $0.595{\pm}0.112$ & $0.667{\pm}0.006$ \\
CrystAF-UMA & $0.844{\pm}0.022$ & $0.490{\pm}0.026$ & $0.812{\pm}0.055$ & $0.625{\pm}0.152$ & $0.437{\pm}0.074$ & $0.635{\pm}0.006$ \\
CrystAF-UMA + chain & $0.829{\pm}0.020$ & $0.522{\pm}0.029$ & $0.773{\pm}0.080$ & $0.501{\pm}0.116$ & $0.447{\pm}0.066$ & $0.664{\pm}0.006$ \\
\bottomrule
\end{tabular}}
\end{table}

\begin{table}[H]
\caption{Structure recovery at $n_s=k=30$ (bootstrap mean $\pm$ SE over samples).}
\label{tab:solved-ns30}
\centering
\small
\setlength{\tabcolsep}{3.5pt}
\resizebox{\textwidth}{!}{%
\begin{tabular}{lcccccc}
\toprule
& Rigid (50) & Flexible (50) & CSP5 (6) & CSP6 (5) & CSP7 (8) & Teaching (779) \\
\midrule
Clari-M & $0.670{\pm}0.026$ & $0.231{\pm}0.030$ & $0.333{\pm}0.003$ & $0.326{\pm}0.097$ & $0.199{\pm}0.065$ & $0.397{\pm}0.007$ \\
CrystAF-base & $0.643{\pm}0.032$ & $0.232{\pm}0.031$ & $0.578{\pm}0.103$ & $0.200{\pm}0.000$ & $0.189{\pm}0.072$ & $0.423{\pm}0.008$ \\
CrystAF-base + chain & $0.640{\pm}0.029$ & $0.239{\pm}0.028$ & $0.459{\pm}0.078$ & $0.200{\pm}0.000$ & $0.202{\pm}0.062$ & $0.411{\pm}0.007$ \\
CrystAF-UMA & $0.640{\pm}0.018$ & $0.180{\pm}0.024$ & $0.649{\pm}0.143$ & $0.327{\pm}0.096$ & $0.204{\pm}0.061$ & $0.427{\pm}0.007$ \\
CrystAF-UMA + chain & $0.624{\pm}0.028$ & $0.262{\pm}0.033$ & $0.477{\pm}0.059$ & $0.200{\pm}0.000$ & $0.229{\pm}0.048$ & $0.423{\pm}0.007$ \\
\bottomrule
\end{tabular}}
\end{table}

\paragraph{Why Clari-M uses 39 NFE here.} Table~\ref{tab:solved} follows Clari's own structure-recovery protocol, which samples with 20 Heun steps (39 NFE); Table~\ref{tab:main} uses Clari's Table-1 protocol with 50 steps (99 NFE). Every row of Table~\ref{tab:solved} uses the same targets, candidate counts, UMA-energy ranking, and COMPACK settings; each model keeps its own sampler.

\subsection{Experimental-structure validation of UMA ranking}
\label{sec:external-validation}

On all 119 OXtal targets, we rank 400 candidates by UMA energy and evaluate the top 200 against the deposited structure with COMPACK. A success is collision-free, matches at least 8 of 15 molecules, and has RMSD below $2$~\AA. Within targets containing both successes and failures, CrystAF-UMA gives an energy-enrichment AUC of $0.549$ (target-bootstrap 95\% CI $[0.520,0.577]$), versus $0.523$ for Clari-M and $0.536$ for Clari-L; its best COMPACK match has mean energy rank $27.6/200$. UMA ranking therefore preferentially retrieves experimentally observed packings. The accompanying repository provides the complete per-target analysis.

\subsection{Representative generated crystals}
\label{sec:qualitative}

\begin{figure}[H]
\centering
\includegraphics[width=\textwidth]{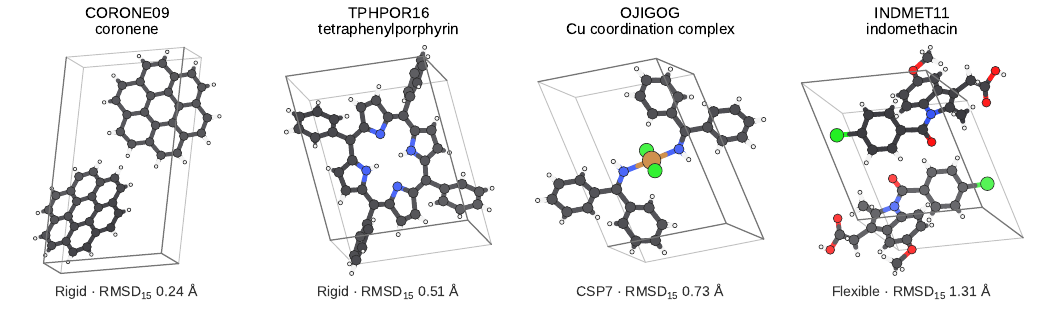}
\caption{Representative CrystAF-generated crystals and COMPACK matches. Each panel shows the best-matched candidate among $400$ samples at $\NFE{=}32$ without sampling-time corrections, with benchmark subset and $15$-molecule COMPACK RMSD.}
\label{fig:cases}
\end{figure}

Figure~\ref{fig:cases} shows examples from one 32-NFE CrystAF-UMA model, without any correction. The best UMA-ranked candidate among 400 samples is shown for a planar aromatic (CORONE09, RMSD$_{15}=0.24$\,\AA), tetraphenylporphyrin (TPHPOR16, $0.51$\,\AA), a Cu coordination complex from CSP7 (OJIGOG, $0.73$\,\AA), and a flexible drug (INDMET11, $1.31$\,\AA). The same weights therefore recover recognisable packings across planar aromatics, large conjugated molecules, coordination chemistry, and flexible pharmaceuticals.

\subsection{Additional negative results and limitations}
\label{sec:limits}

\paragraph{Second-seed variation.}
A second seed reproduces the main validity gain of the PoseBusters-ranked reward but not its best clash and volume numbers: the two seeds give $93.21/7.86/1.61$ and $92.56/8.60/1.88$ for PB/clash/volume. The physical-reward seeds give $93.01/7.98/1.88$ and $93.48/8.60/2.01$. The reward families therefore overlap on the headline validity result, while the single best clash and volume cells are seed-sensitive.

\paragraph{Relax-and-distill.}
See Section~\ref{sec:main}; relaxing the cell in the targets makes the volume bias worse ($+4.16\%\to+6.50\%$ on 36 structures), so the targets keep the generated cell.

\paragraph{Scaling the reward.}
Tripling families per epoch and increasing candidates per family from 100 to 150 tracks the same quality curve as the matched 80-family control when compared at the same number of families seen, then collapses in the eighth epoch. Additional sampling therefore reduces wall-clock through parallelism but does not raise the quality ceiling in these runs.

\paragraph{Corrector trade-off.}
The sampling-time correction chain strongly improves clash and stereochemistry but modestly worsens PDD. This is why the uncorrected CrystAF-UMA row remains the primary evidence for learned physical alignment and the corrected rows are reported as optional operating points.

\end{document}

%% file: math_commands.tex
\usepackage{amsmath,amsfonts,bm}

\def\eqref#1{equation~\ref{#1}}
\def\1{\bm{1}}

\DeclareMathAlphabet{\mathsfit}{\encodingdefault}{\sfdefault}{m}{sl}
\SetMathAlphabet{\mathsfit}{bold}{\encodingdefault}{\sfdefault}{bx}{n}